\documentclass[modern]{aastex631}

\usepackage{amsmath}
\usepackage{amssymb}
\usepackage{hyperref}
\usepackage{bm}
\usepackage{listings}
\usepackage{graphicx}
\usepackage{float}
\usepackage{soul}
\usepackage{fontawesome}

\newcommand{\nuancemethod}{\textit{nuance}}
\newcommand{\nuance}{\nuancemethod{}}
\newcommand{\nuancecode}{\textsf{nuance}}

\newcommand{\review}[1]{#1}
\newcommand{\set}[1]{\{\,#1\,\}}
\newcommand{\footlink}[1]{\footnote{\url{#1}}}
\newcommand{\wtls}{\texttt{bi-weight+BLS}}
\DeclareMathOperator*{\argmax}{arg\,max}

\lstdefinestyle{mystyle}{
    backgroundcolor=\color{white},
    commentstyle=\color{gray!70},
    keywordstyle=\color{Bittersweet},
    stringstyle=\color{RoyalBlue},
    basicstyle=\fontsize{8.5}{13}\fontfamily{DejaVuSansMono-TLF}\selectfont,
    breakatwhitespace=false,         
    breaklines=true,
    rulecolor=\color{black!15},
    numbers=none,
    numberstyle=\fontsize{7}{11}\fontfamily{DejaVuSansMono-TLF}\selectfont\color{gray!50},
    framerule=0pt,
    breakindent=5pt,
    resetmargins=true,
    numbersep=10pt,
    frame=single,
    aboveskip=1em,
    belowskip=1em,
    xleftmargin=6pt,
    framexleftmargin=4pt
}
\definecolor{linkcolor}{rgb}{0.1216,0.4667,0.7059}
\newcommand{\codeicon}{{\color{linkcolor}\faFileCodeO}}
\newcommand{\codelink}[1]{\href{https://github.com/lgrcia/paper-nuance/blob/main/#1}{\codeicon}\,\,}
\newcommand{\outcodelink}[1]{\href{#1}{\codeicon}\,\,}
\newcommand{\gitlink}[1]{\href{#1}{\color{linkcolor}\faGithub}\,\,}

\graphicspath{{./figures}}

\begin{document}

\title{\texttt{nuance}: Efficient detection of planets transiting active stars}

% author info
\author[0000-0002-4296-2246]{Lionel J. Garcia}
\affiliation{Center for Computational Astrophysics, Flatiron Institute, New York, NY, USA}
\affiliation{\review{Astrobiology Research Unit, University of Liège, Liège, Belgium}{}}
\author[0000-0002-9328-5652]{Daniel Foreman-Mackey}
\affiliation{Center for Computational Astrophysics, Flatiron Institute, New York, NY, USA}
\author[0000-0001-8504-5862]{Catriona A. Murray}
\affiliation{Department of Astrophysical and Planetary Sciences, University of Colorado, Boulder, USA}
\author[0000-0003-1453-0574]{Suzanne Aigrain}
\affiliation{Sub-department of Astrophysics, Department of Physics, University of Oxford, Oxford, UK}
\author[0000-0002-2457-7889]{Dax L. Feliz}
\affiliation{American Museum of Natural History, New York, NY, USA}
\author[0000-0003-1572-7707]{Francisco J. Pozuelos}
\affiliation{Instituto de Astrofísica de Andalucía, Granada, Spain}
\affiliation{Astrobiology Research Unit, University of Liège, Liège, Belgium}
\correspondingauthor{Lionel J. Garcia}
\email{lgarcia@flatironinstitute.org}

\keywords{exoplanet detection methods, stellar activity, time series analysis, gaussian processes regression, computational methods, GPU computing}

\begin{abstract}
    The detection of planetary transits in the light curves of active stars, featuring correlated noise in the form of stellar variability, remains a challenge. Depending on the noise characteristics, we show that the traditional technique that consists of detrending a light curve before searching for transits alters their signal-to-noise ratio, and hinders our capability to discover exoplanets transiting rapidly-rotating active stars. We present \nuancemethod{}, an algorithm to search for transits in light curves while simultaneously accounting for the presence of correlated noise, such as stellar variability and instrumental signals. We assess the performance of \nuancemethod{} on simulated light curves as well as on the TESS light curves of 438 rapidly-rotating M dwarfs. For each dataset, we compare our method to 5 commonly-used detrending techniques followed by a search with the Box-Least-Squares algorithm. Overall, we demonstrate that \nuancemethod{} is the most performant method in 93\% of cases, leading to both the highest number of true positives and the lowest number of false positive detections. Although simultaneously searching for transits while modeling correlated noise is expected to be computationally expensive, we make our algorithm tractable and available as the \textsf{JAX}-powered Python package \href{https://github.com/lgrcia/nuance}{\nuancecode{}}, allowing its use on distributed environments and GPU devices. Finally, we explore the prospects offered by the \nuancemethod{} formalism, and its use to advance our knowledge of planetary systems around active stars, both using space-based surveys and sparse ground-based observations. \href{https://github.com/lgrcia/nuance}{\color{linkcolor}\faCode}\,\,\href{https://github.com/lgrcia/paper-nuance}{\color{linkcolor}\faFileTextO}\,\,\href{https://nuance.readthedocs.io}{\color{linkcolor}\faBook}
\end{abstract}

\section*{Introduction}
Transiting exoplanets are keystone objects for the field of exoplanetary science, but detecting transits in light curves featuring stellar variability and instrumental signals remains a challenge (e.g. \citealt{Pont2006}, \citealt{Howell2016} or \citealt{Yaptangco2024}). For this reason, known transiting exoplanets tend to be found around quieter stars, or belong to the population of close-in giants whose transit signals dominate over stellar rotational variability \citep{Simpson2023}. However, transiting exoplanets around active stars are discoveries with significant scientific value. First, as younger stars are more active \citep{Skumanich1972}, being able to detect planets transiting active stars will favor the discovery of young planetary systems \citep[e.g.][]{Newton2022}. Second, as stellar variability may originate from surface active regions (such as starspots), transiting exoplanets can be used to map the photosphere of active stars \citep[e.g.][]{Morris2017}, benefiting both the study of stellar atmospheres and the concerning impact of their nonuniformity on planetary atmosphere retrievals \citep{rackham2018}. Overall, enabling the detection of transits in light curves with high levels of correlated noises will greatly benefit the study of terrestrial exoplanets around late M-dwarfs, usually observed at lower SNR and more likely to display photometric variability (e.g. \citealt{Murray2020} or \citealt{Petrucci2024}).
\\\\
Commonly used transit-search algorithms, such as the Box-Least-Squares algorithm \citep[BLS,][]{bls} are capable of detecting transits in light curves containing only transit signals and white noise. Using this method, the simplest way to detect transits in a light curve featuring correlated noise (either astrophysical or instrumental), is to first clean it from nuisance signals before performing the search. This strategy is widely adopted by the community, both using physically-motivated systematic models like \cite{everest1, everest2}, or filtering techniques (\citealt{Jenkins2010}, \citealt{wotan}). However, when correlated noise starts resembling transits, this cleaning step, often called \textit{detrending}, is believed to degrade their detectability \cite[see subsection 4.3 of][]{wotan}. In this case, the only alternative to search for transits is to perform a full-fledged modeling of the light curve, including both transits and correlated noise, and to compute the likelihood of the data to the transit model on a wide parameter space, an approach largely avoided due to its intractable nature. Nonetheless, \cite{kovacs2016} ask: \textit{Periodic transit and variability search with simultaneous systematics filtering: Is it worth it?}. The latter study explores a handful of cases and generally discards the benefit of using a full-fledged approach.\\\\
\newpage
In this paper, we identify regions of light curves morphological parameter space for which a full-fledged transit search is necessary, and we present \nuancemethod{}\footnote{Throughout the paper, \nuancemethod{} written in italics refers to the algorithm, while \nuancecode{} in sans-serif refers to its implementation published as a Python package in its version 0.5.2 (see \autoref{package}).}, a method to search for transit signals while simultaneously modeling correlated noises in a tractable way. In \autoref{issues}, we describe the effect of correlated noise on transit light curves and the effect of its detrending on transit signals detectability. In \autoref{nuance}, we present \nuance{}, and the two main steps on which this method is based: the linear search and the periodic search. In \autoref{results}, we test the performance of \nuance{} on a wide variety of cases, and compare our method to commonly used transit search algorithms. This include transits injected in synthetic datasets, but also in the TESS light curves of 438 rapidly-rotating M dwarfs. Finally, in \autoref{discussion}, we discuss the results and the limitations of \nuance{}, before concluding in \autoref{conclusion}.\\\\
This paper was compiled from source code available in a git-versioned repository (\href{https://github.com/lgrcia/paper-nuance}{\color{linkcolor}\faFileTextO}) and used the \textsf{snakemake}\footlink{https://snakemake.readthedocs.io} workflow management tool. Source code of the \textsf{nuance} \textsf{Python} package is also available in a git-versioned repository (\href{https://github.com/lgrcia/nuance}{\color{linkcolor}\faCode}) together with an online documentation (\href{https://nuance.readthedocs.io}{\color{linkcolor}\faBook}). Finally, we provide links to the source codes used to produce each figure (indicated by {\color{linkcolor}\codeicon}).

\newpage
\section{Motivation}\label{issues}
A strong assumption when using the BLS algorithm to search for transits is that the searched dataset only contains transit signals and white noise, justifying the need for detrending. In this section, we explore this effect by simulating light curves containing a transit signal and correlated noise in the form of stellar variability, and study how detrending the latter affects the transit signal detectability depending on the light curve morphological characteristics. Hence, we first present how transit light curves are simulated, using a stochastic model of stellar variability, and describe the detection metric we employ to quantify transit detectability in the presence of correlated noise.
\subsection{Light curve simulations}\label{light_curves_simulations}
\begin{figure}[H]
    \begin{centering}
        \includegraphics[width=\linewidth]{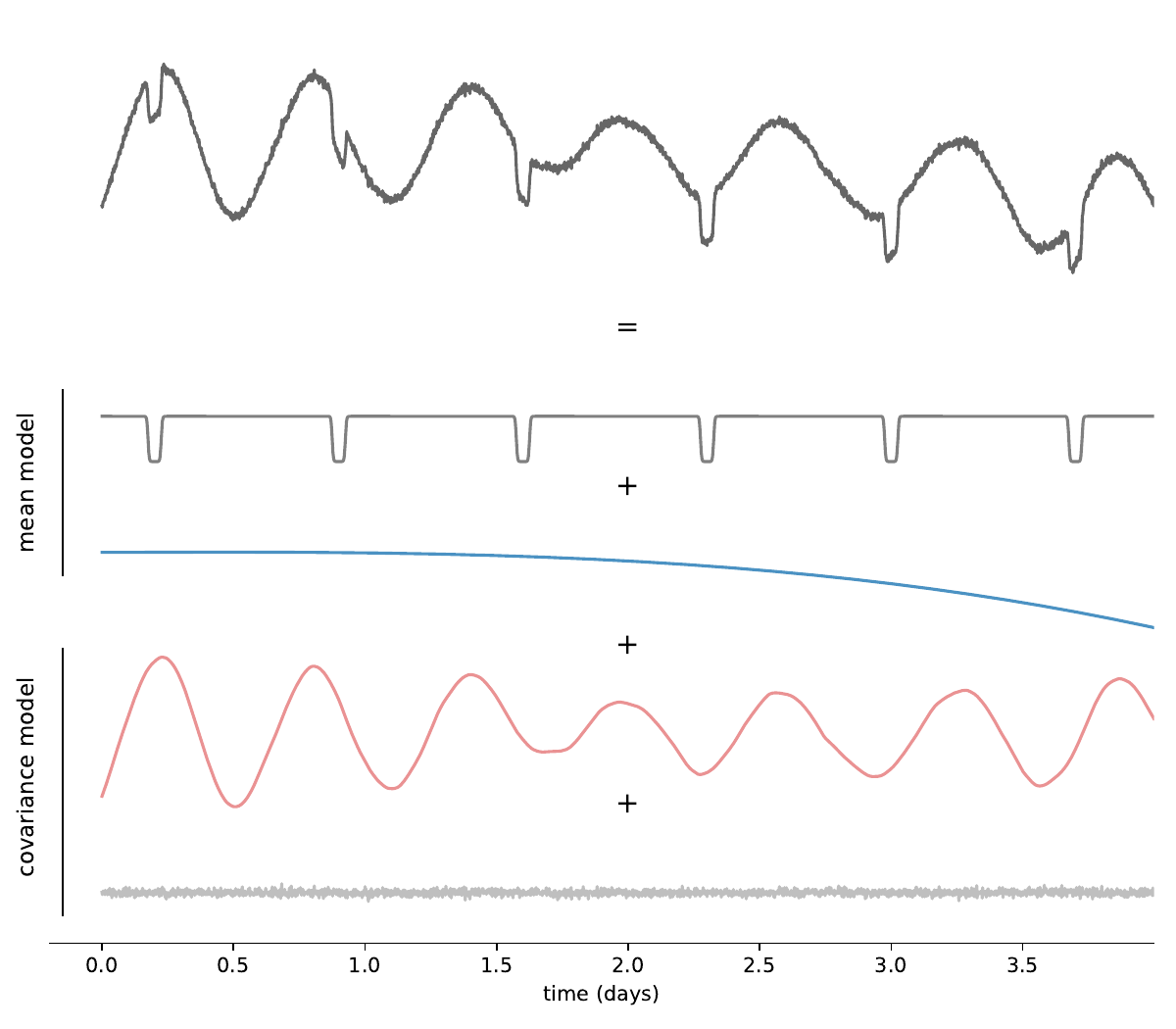}
        \caption{The top plot shows the simulated flux time series of an active star drawn from the GP model described in \autoref{light_curves_simulations}. This dataset corresponds to an observation of 4 days with an exposure time of 2 minutes. The mean of this signal consists of a periodic transit signal of period $P=0.7$ days, duration $D=1.2$ hours and depth $\Delta=2\%$ (dark gray) plus an instrumental signal (blue). Correlated noise in the form of stellar variability is simulated with an SHO kernel of hyperparameters $\omega = \pi / 6 D$, $Q=45$ and $\sigma = \Delta$. Finally, white noise with a standard deviation of $0.1\%$ is added to the diagonal of the covariance matrix. \codelink{workflows/principle/principle.ipynb}}
        \label{fig:app_principle_dataset}
    \end{centering}
\end{figure}

To perform this study, we want to simulate realistic light curves that contain instrumental signals, transit signals and correlated noise with characteristics that can be controlled using a set of interpretable parameters. To this end, we model light curves as realizations of a Gaussian process (GP; \citealt{Rasmussen2005}, \citealt{Aigrain2023}) with a mean containing the instrumental and transit signals, and a kernel allowing to model different forms of correlated noise controlled by its hyperparameters.\\\\
Let $f$ be the simulated differential flux of a star sampled and arranged in the vector $\bm{f}$ associated to the vector of times $\bm{t}$, such that
\begin{equation*}
    \bm{f} \sim \mathcal{N}(\bm{\mu}, \bm{\Sigma}),
\end{equation*}
i.e.\;that $\bm{f}$ is drawn from a GP of mean $\bm{\mu}$ and covariance matrix $\bm{\Sigma}$. For practical reasons, we model $\bm{\mu}$ as a linear combination of $M$ explanatory variables, such that
\begin{equation}\label{eq:linear_model}
    \bm{\mu} = \bm{X w}
\end{equation}
where the first $M-1$ columns of $\bm{X}$ are contemporaneous instrumental time series measurements, such as the position of the star on the detector, the sky background, co-trending basis vectors, or any other explanatory variables. The last column of $\bm{X}$ is a box-shaped transit signal with a fixed epoch, duration and period. This way, the transit signal is part of the mean linear model, meaning that once the design matrix $\bm{X}$ is constructed the transit is only parametrized by its depth $\Delta$.\\\\
As we are interested in active stars whose fluxes feature correlated noise in the form of stellar variability, we choose a physically-motivated GP kernel to model the covariance matrix $\bm{\Sigma}$, describing stellar variability through the covariance of a stochastically-driven damped harmonic oscillator (SHO, \citealt{celerite, celerite2}) parametrized by its quality factor $Q$, its pulsation $\omega$ and the amplitude of the kernel function $\sigma$ (the full expression of the kernel function is provided in \autoref{app_gp}). This choice of mean and kernel function completely defines the GP, from which light curves with different levels of correlated noise can be drawn (see an example in \autoref{fig:app_principle_dataset}).
\subsection{Transit detectability}\label{transit_detectability}

One way to quantify the detectability of a unique transit signal is to compute its signal-to-noise-ratio expressed as 
\begin{equation*}
  SNR= \frac{\Delta}{\sigma}\sqrt{n},
\end{equation*}
where $\Delta$ is the transit depth, $\sigma$ the measurements uncertainty (assuming homoscedasticity), and $n$ the number of points within transit. Although this metric is useful to assess the strength of the transit signal given a certain photometric precision, it does not account for the presence of correlated noise. However, instrumental and other astrophysical signals will necessarily affect the detectability of transits in realistic light curves \citep{Pont2006}. For this reason, the combined differential photometric precision (CDPP; \citealt{Jenkins2010}) metric was developed, and used in the context of the Kepler mission to assess the level of correlated noise in light curves, affecting the detectability of transits with a given duration. 
% Ultimately, this metric was essential to estimate the yield of the survey, and helped to achieve one of the primary goal of the mission: to measure the occurrence rate of Earth-size planets orbiting in the habitable zones of sun-like stars \citep{}.
The CDPP is computed by decomposing the data in the time-frequency domain using wavelets, and measures the significance of a distorted transit signal in the whitened data. For a given transit duration, the CDPP is then a measure of the noise remaining after filtering the light curve in each frequency band, taking into account the presence of nonstationary correlated noise of a given timescale (see \citealt{Jenkins2010} for more details). Hence, we can estimate the significance of the transit signal accounting for the presence of correlated noise as
\begin{equation}\label{eq:snr}
  SNR= \frac{\hat{\Delta}}{CDPP_D},
\end{equation}
where $CDPP_D$ is the CDPP computed for a given transit duration $D$ and $\hat{\Delta}$ is the transit depth after detrending. For simplicity, the CDPP is computed using the method from \cite{Gilliland2011} implemented in the \textsf{lightkurve} Python package \citep{Lightkurve} and $\hat{\Delta}$ is the depth obtained by solving \autoref{eq:linear_model} with the last column of $\bm{X}$ containing a normalized transit model.
\begin{figure}[H]
    \begin{centering}
        \includegraphics[width=0.6\linewidth]{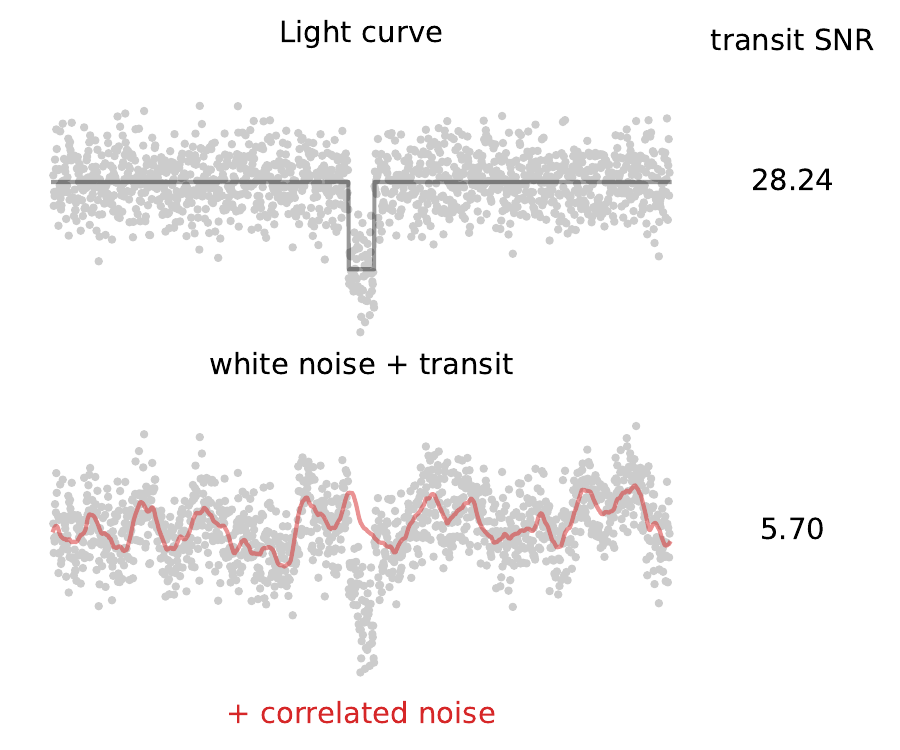}
        \caption{Illustration of the effect of correlated noise on a single transit SNR. An hour-long transit of depth 0.5\% is generated on top of white noise (standard deviation of 0.15\%) as part of a 24-hours observation with an exposure time of 1 minute (top). Then, in the bottom plot, correlated noise is added, generated using a GP with an SHO kernel of hyperparameters $\omega=60$, $Q=0.5$ and $\sigma=\Delta/4$ . The SNR on the right of each light curve is computed using \autoref{eq:snr}. Models used to simulate these data are provided in \autoref{light_curves_simulations}. \codelink{workflows/plot_issues/scripts/plot_issue_1.py}}
        \label{fig:issue1}
    \end{centering}
\end{figure}
\autoref{fig:issue1} shows the SNR from \autoref{eq:snr} computed for a unique transit observed in the absence (gray) and presence of correlated noise (red). As illustrated in this figure, the presence of correlated noise strongly decreases the transit signal SNR, which would ultimately limit its detectability.

\subsection{Detrending methods and their effects}\label{detrending_effect}
\begin{figure}[H]
    \begin{centering}
        \includegraphics[width=\linewidth]{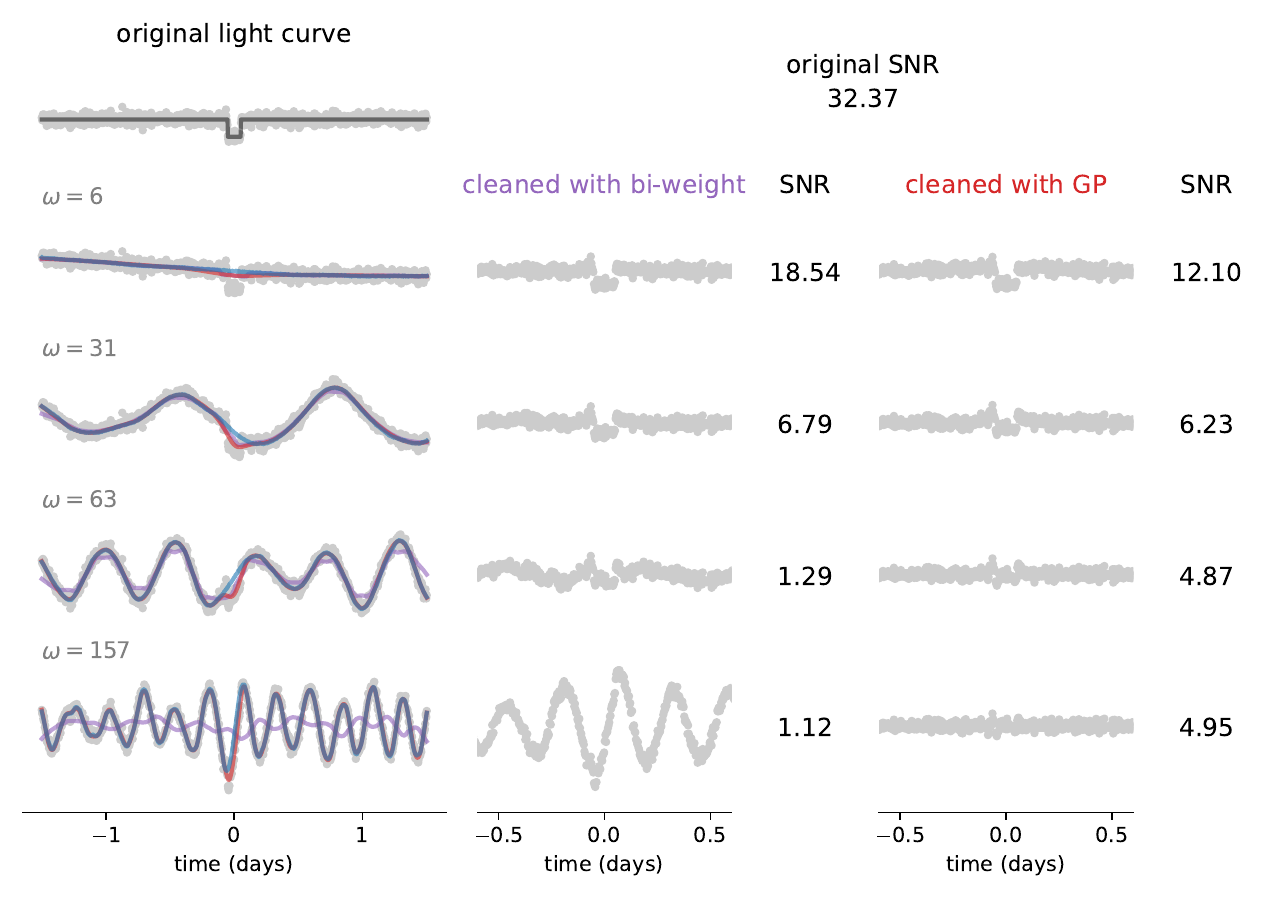}
        \caption{In each plot on the left, correlated noises simulated using a GP with an SHO kernel are added (blue line) to an original light curve containing a transit of depth $\Delta = 0.8\%$, duration $D=0.1$ day, and white noise with a standard deviation of 0.15\% (top left). The hyperparameters of the kernel are fixed to $Q=10$ and $\sigma=0.8\%$, with the pulsation $\omega$ increasing from top to bottom. The resulting stellar variability signals are then reconstructed using (in purple) Tukey's bi-weight filter with an optimal window size of $3\times D$  (see \citealt{wotan}), and (in red) a GP with the same kernel used to simulate the data. In each case the variability is reconstructed, subtracted, and the transit SNR computed using \autoref{eq:snr}. \codelink{workflows/plot_issues/scripts/plot_issue_2.py}}
        \label{fig:issue2}
    \end{centering}
\end{figure}

The presence of instrumental correlated noise motivated the development of systematics detrending algorithms, such as the Trend Filtering Algorithm (\textsc{TFA}, \citealt{tfa}, in its primary use case), \textsc{SysRem} (\citealt{sysrem}) or Pixel Level Decorrelation (\textsc{PLD}, \citealt{pld}; see also \textsc{Everest} from \citealt{everest1, everest2}). Most of these methods rely on the shared nature of instrumental signals among light curves (or neighboring pixels) such that the correction applied should not degrade the transit signal and can be modeled using contemporaneous measurements (e.g.\;detector's temperature, pointing error, sky background or airmass time series). But even after instrumental signals have been removed, stellar variability and other astrophysical signals remain, which gave rise to several approaches. Some of them are physically-motivated and make use of GPs (e.g.\;\citealt{k2sc}), others are empirical and make use of filtering and data-driven algorithms (\citealt{Jenkins2010}, \citealt{wotan}). In this section, we show how these techniques impact transits detectability, depending on the morphological characteristics of light curves.\\\\
In \autoref{fig:issue2}, we simulate a transit signal on top of which we add photometric stellar variability with different timescales, sampled from a GP with an SHO kernel described in \autoref{app_gp}. For each light curve, we reconstruct and detrend stellar variability in two ways: one using the widely-adopted Tukey's bi-weight filter, presented in \cite{tukey} and using the implementation from \textsf{wõtan}\footnote{\href{https://github.com/hippke/wotan}{https://github.com/hippke/wotan}} \citep{wotan}; the other using the same GP from which the data have been sampled. We then estimate the resulting transit depth and compute the remaining transit SNR using \autoref{eq:snr}. \autoref{fig:issue2} clearly shows the effect of both detrending techniques on transits SNR, and intuitively suggests that this degradation due to detrending is strongly dependent on the correlated noise characteristics encountered.
To explore the parameter space for which detrending is the most problematic, we employ the model described in \autoref{light_curves_simulations} to simulate 10 000 differential light curves with different morphological characteristics, and compute the remaining transit SNR after detrending. In order to place the stellar variability hyperparameters on a relative scale with the transit signal parameters, we reparametrize the SHO kernel with
\begin{equation}\label{eq:relative_params}
    \tau = \frac{\pi}{\omega D}, \hspace{0.5cm} 
    \delta = \frac{2\sigma}{\Delta} \hspace{0.5cm}  \text{and}  \hspace{0.5cm}  
    Q = 10,
\end{equation}
where $\tau$ is the relative timescale of the variability with respect to the transit duration and $\delta$ the relative amplitude of the variability against the transit depth, both being adimensional. Hence, for $(\tau, \delta)=(1, 1)$, the expressions of $\omega$ and $\sigma$ given in \autoref{eq:relative_params} correspond to a variability signal with a period half that of the transit duration, and a correlated noise amplitude comparable to the transit depth, i.e. strongly resembling the simulated transit signal.\\\\
For each of the 10 000 light curves generated, we separately reconstruct and detrend the variability signal using the two techniques employed in \autoref{fig:issue2}, i.e.\;one using an optimal bi-weight filter with a window size three times that of the transit duration \citep{wotan} and the other using a GP with an optimal kernel. We then compute the detrended transit SNR using \autoref{eq:snr}.
\begin{figure}[H]
    \begin{centering}
        \includegraphics[width=\linewidth]{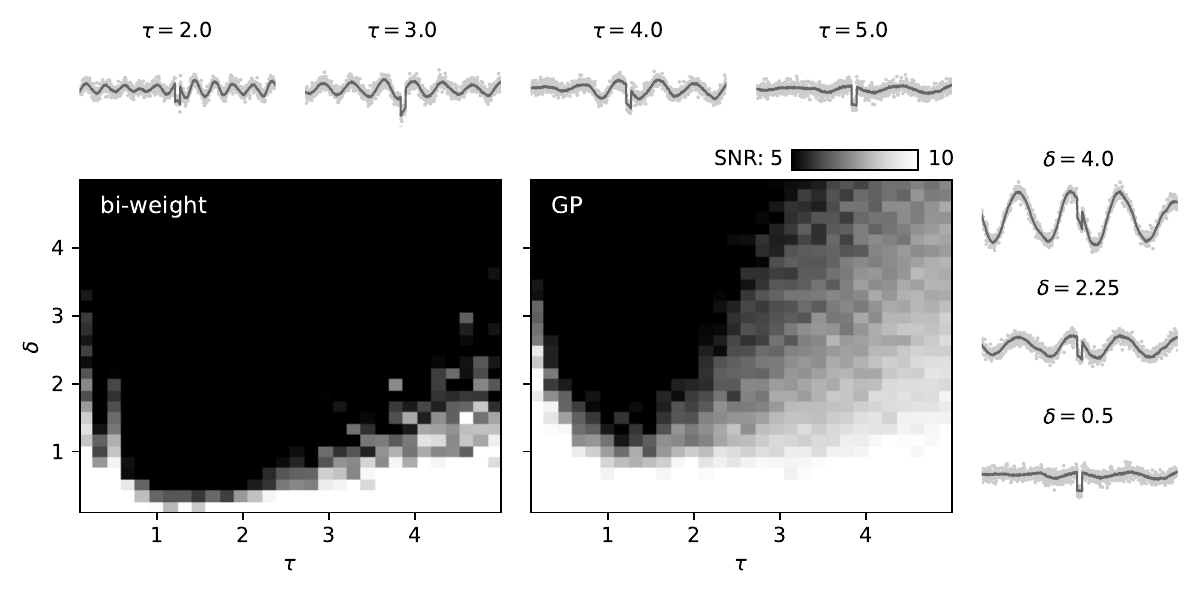}
        \caption{Transit SNR computed for 10 000 differential light curves with varying morphological characteristics. All light curves span 2.8 days with an exposure time of 2 minutes and contain the same transit signal of  duration $D=1$ hour, depth of 0.5\% and white noise with a standard deviation of 0.1\%. Light curves at the top and right side of the central plot are shown with their corresponding $\tau$ and $\delta$ values, which again correspond to the relative timescale of stellar variability against the transit duration, and the relative amplitude of stellar variability against the transit depth. The two central plots show the transit SNR after detrending the light curves with one of two methods. On the left, light curves are detrended using Tukey's bi-weight filter with an optimal window size of $3\times D$ (see \citealt{wotan}), while on the right, light curves are detrended using a GP with the same kernel used to simulate the data. \codelink{workflows/cleaning_snr/scripts/plot.py}}
        \label{fig:snr_detrend}
    \end{centering}
\end{figure}
\autoref{fig:snr_detrend} shows that there exists an entire region in the $(\tau, \delta)$ parameter space for which the bi-weight detrending degrades transit SNR to the point of no detection ($SNR < 5$). Although being more robust, the same effect is observed when detrending with an optimal GP. Hence, detrending makes transit search blind to many systems, especially when the widely adopted bi-weight filter is used. We note that, outside this problematic parameter space, both techniques perform relatively well and such a degradation of the transit SNR should not be expected. Although this study should be extended to other detrending techniques, it highlights the need for a more informed transit search algorithm able to deal with correlated noise, at least if present in the form of stellar variability.

\newpage
\section{\textsf{nuance}}\label{nuance}
When searching for exoplanets using an indirect method, such as using transits or radial velocities, the initial detection is often as important as the follow-up observations, ultimately leading to the confirmation of the planetary candidate. For that reason, the final product of most transit search algorithms consists of a \textit{periodogram}: a 1D detection metric given a set of trial periods that allows to identify the presence of a potential candidate, and the period and epoch at which it should be followed up.\\\\ 
If we assume that a transit is defined by its period $P$, epoch $T_0$, duration $D$ and depth $\Delta$, we then wish to compute the likelihood of such a transit being present in the data for a set of periods $P$, leading to an interpretable periodogram. As we are interested in following up transits with specific parameters, such as a well-defined epoch, the likelihood we wish to compute must not be marginalized over all parameters other than $P$ but rather specifically computed at the maximum-likelihood parameters $\hat T_0$, $\hat D$ and $\hat \Delta$, leading to the \textit{profile likelihood}
\begin{equation}\label{eq:periodogram}
       \mathcal{Q}(P) = p(\bm{f} \vert P, \hat T_0 ,\hat D, \hat \Delta),
\end{equation} 
where $\bm{f}$ denotes the data. In this section, we present \textit{nuance}, an algorithm to compute transit-search periodograms in a tractable way, and detect planetary transits in light curves containing correlated noise such as instrumental signals and photometric stellar variability. In \autoref{approach}, we explain how our approach requires two separate steps in order to remain tractable, and present these steps in \autoref{linear_search} and \autoref{periodic_search}, leading to the transit-search periodogram described in \autoref{periodogram}. Finally, we present the \nuancecode{} Python package in \autoref{package} and perform a control test of our implementation against BLS in \autoref{control}.

\subsection{The approach}\label{approach}
As in \autoref{light_curves_simulations}, let assume that $\bm{f}$ is a vector of size $N$ containing the flux of a star observed at times $\bm{t}$ such that
\begin{equation*}
    \bm{f} \sim \mathcal{N}(\bm{X w}, \bm{\Sigma}),
\end{equation*}
i.e.\;that $\bm{f}$ is drawn from a GP of mean $\bm{Xw}$ and covariance $\bm{\Sigma}$. Again, we set the first $M-1$ columns of the $(N\times M)$ design matrix $\bm{X}$ as contemporaneous measurements and the last column as a normalized box-shaped transit of period $P$, epoch $T_0$ and duration $D$. This way, the transit signal is part of the linear model and its depth $\Delta$ can be solved linearly. Under this assumption, the log-likelihood of the data given the presence of a periodic transit signal of period $P$, epoch $T_0$, duration $D$, and a mean linear model with coefficients $\bm{w}$ is \citep{Rasmussen2005}
\begin{equation} \label{eq:linear_search_ll}
    \ln p(\bm{f} \vert P, T_0, D, \bm{w}) = -\frac{1}{2}(\bm{f}-\bm{Xw})^T\bm{\Sigma}^{-1}(\bm{f}-\bm{Xw}) -  \frac{1}{2}\ln\vert\bm{\Sigma}\vert - \frac{N}{2}\ln 2\pi.
\end{equation}
Given the linearity of the mean model at $P$, $T_0$ and $D$ fixed, this likelihood is maximized for the least-square parameters
\begin{equation}\label{eq:ls}
    \bm{w} = (\bm{X}^T\bm{\Sigma}^{-1}\bm{X})^{-1}\bm{X}^T\bm{\Sigma}^{-1}\bm{f} \hspace{0.5cm} \text{with  uncertainties} \hspace{0.5cm} \bm{\sigma} = (\bm{X}^T\bm{\Sigma}^{-1}\bm{X})^{-1}.
\end{equation}
As we are only interested in the transit depth, we will omit $\bm{w}$ in the remainder of this paper and simply write likelihood expressions depending on its last value $\Delta=\bm{w}_M$, such as $p(\bm{f} \vert P, T_
0, D, \Delta)$, where all other parameters of $\bm{w}$ are taken at their maximum-likelihood values. \\\\
Hence, given a period $P$, computing $\mathcal{Q}(P)$ boils down to a nonlinear optimization involving several evaluations of the likelihood given in \autoref{eq:linear_search_ll}. While this is tractable for few periods, it becomes highly untractable for the large set of trial periods required to properly sample a transit-search periodogram (in the order of tens of thousands).\\\\
To remain tractable, \nuancemethod{} employs an extension of the strategy used by \cite{foreman2016}, which has significant intellectual overlap with the methods used by \cite{Aigrain2004} and \cite{Jenkins2010}. This approach separates the transit search into two components: the linear search and the periodic search. During the linear search, the likelihood $p(\bm{f} \vert T_0, D, \Delta)$ of a single nonperiodic transit is computed for a grid of epochs and durations, each time solving for $\Delta$ linearly. Then, the periodic search consists of combining these likelihoods to compute the likelihood $p(\bm{f} \vert P, T_0, D, \Delta)$ of the data given a periodic transit signal for a range of periods $P$. These combined likelihoods yield $p(\bm{f} \vert P, \hat T_0 ,\hat D, \hat \Delta)$, a transit-search periodogram on which the periodic transit detection is based. \nuancemethod{} differs from \cite{foreman2016} and other existing transit search algorithms as it models the covariance of the light curve with a GP, accounting for correlated noise (especially in the form of stellar variability) while keeping the model linear and tractable. This way, \nuancemethod{} searches for transits while, at the same time, modeling correlated noise, avoiding the separate detrending step that degrades transit signals SNR (see \autoref{issues}).\\\\
% We note that the approach employed by \nuancemethod{} (using the two steps proposed by \citealt{foreman2016}), shares similarities with the approach of \cite{Jenkins2010}, where a single event statistic is computed and combined into a multiple event statistics.

\subsection{The linear search}\label{linear_search}

\begin{figure}[H]
    \begin{centering}
        \includegraphics[width=\linewidth]{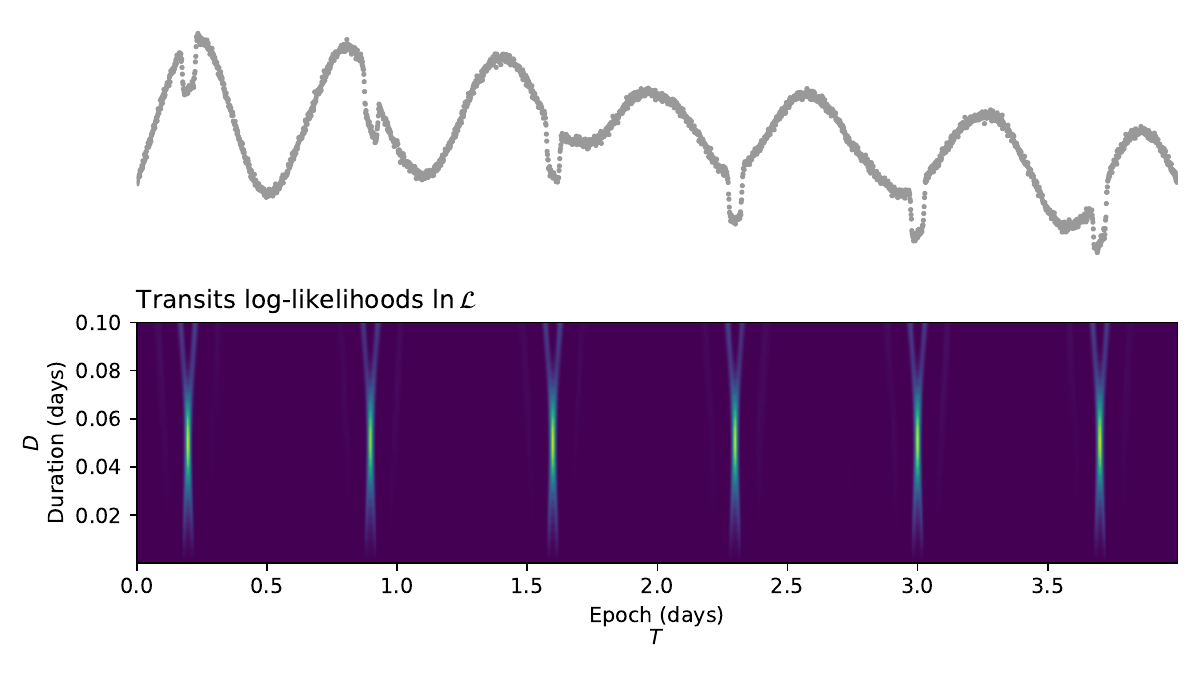}
        \caption{Principle and output of the linear search. The simulated dataset (top) corresponds to the one shown and described in \autoref{fig:app_principle_dataset}. First, a set of durations and depths $\set{T_i, D_j}_{i,j}$ is generated. For each pair of indices $(i,j)$, the likelihood $\ln p(\bm{f} \vert T_i ,D_j, \Delta_{i,j})$ is computed using the parameters from \autoref{eq:ls} and the expression of \autoref{eq:linear_search_ll}. This process yields the grid of log-likelihoods $\ln\mathcal{L}$ (bottom plot), as well as the $\set{\Delta_{i,j}, \sigma_{i,j}}_{i, j}$ transit depths and errors inferred linearly using \autoref{eq:ls}. \codelink{workflows/principle/principle.ipynb}}
        \label{fig:linear_search}
    \end{centering}
\end{figure}

The goal of the linear search is to compute the likelihood $p(\bm{f} \vert T_0 , D, \Delta)$ for a grid of epochs $\set{T_i}_i$ and durations $\set{D_j}_j$. For each pair $(T_i, D_j)$ the transit depth $\Delta_{i, j}$ is linearly solved, which leads to the set of maximum likelihoods
\begin{equation*}
    \set{\ln\mathcal{L}_{i,j}}_{i, j} = \set{\ln p(\bm{f} \vert T_i ,D_j, \Delta_{i, j})}_{i, j}.
\end{equation*}
An example of such a grid of likelihoods is shown in \autoref{fig:linear_search}. To prepare for the next step, uncertainties on the depths $\sigma_{i,j}$ are also computed using \autoref{eq:ls} and stored.

\subsection{The periodic search}\label{periodic_search}

\begin{figure}[H]
    \centering
    \includegraphics[width=\linewidth]{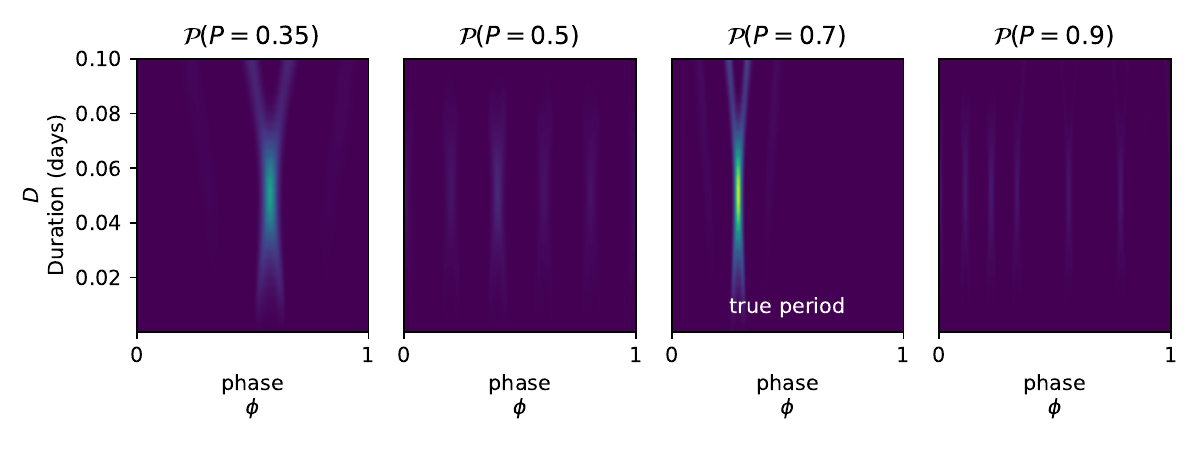}\\
    \caption{Periodic search likelihood $\mathcal{P}(P)$ computed for different trial periods $P$. Notice how the maximum value of $\mathcal{P}$ for the alias period $P=0.35$ day (left plot) is lower than for $P=2\times 0.35=0.7$ day, a result of combining the log-likelihoods using \autoref{eq:result} instead of \autoref{eq:attempt}, in favor of individual transits matching a common depth $\Delta$. \codelink{workflows/principle/principle.ipynb}}
    \label{fig:periodic_search}
\end{figure}

We then need to combine the likelihoods computed from the linear search to obtain
\begin{equation*}
    p(\bm{f} \vert P, T_0 , D, \Delta),
\end{equation*}
i.e.\,the probability of a periodic transit of period $P$, epoch $T_0$, duration $D$ and depth $\Delta$ given the data $\bm{f}$. For a given transit duration $D$, any combination of $(P, T_0)$ leads to K transits, for which it is tempting to write
\begin{equation}\label{eq:attempt}
    p(\bm{f} \vert P, T_0 ,D, \Delta) = \prod_k^K p(\bm{f} \vert T_k, D, \Delta_k),
\end{equation}
where $\set{T_k}_k$ are the epochs matching $(T_0, P)$ and $\set{\Delta_k}_k$ the corresponding depths, so that
\begin{equation*}
    \ln p(\bm{f} \vert P, T_0 ,D, \Delta) = \sum_k^K \ln \mathcal{L}_k.
\end{equation*}
This is the joint likelihood of transits belonging to the same periodic signal but with varying depths  $\set{\Delta_k}_k$. However, individual transits from a periodic signal cannot be considered independent, and should instead be found periodically and share a common transit depth $\Delta$. To this end, it can be shown (see \autoref{proof}) that there is a closed form expression for the joint likelihood of K individual transits with depths and errors $\set{\Delta_k, \sigma_k}_k$ assuming a common depth $\Delta$, corresponding to 
\begin{equation}\label{eq:result}
    \begin{gathered}
        \ln p(\bm{f} \vert P, T_0 ,D, \Delta) =  \sum_{k}^K \ln \mathcal{L}_k  - \frac{1}{2} \sum_k^K\left(\ln(\sigma_{k}^2) - \ln(\sigma^{2} + \sigma_{k}^{2}) +  \frac{\left(\Delta_{k} -
        \Delta\right)^{2}}{\sigma_k^{2} + \sigma^{2}}\right) \\
        \text{with} \quad  \frac{1}{\sigma^2} = \sum_k^K \frac{1}{\sigma_k^2} \quad \text{and} \quad
        \Delta = \sigma^2 \sum_k^K {\frac{\Delta_k}{\sigma_k^2}}.
    \end{gathered}
\end{equation}
In order to compute this joint likelihood, we must assume that the likelihoods computed during the linear search are independent. However, the individual transits combined in the periodic search are not independent. Indeed, by using a GP, we assume that points belonging to separate transits may have a nonzero covariance. In practice, we notice that this covariance is small enough to consider each transit as independent, a reasonable assumption for most physically-realistic datasets.\\\\
While \autoref{eq:result} takes a closed form, the individual epochs matching $T_0$ and $P$ are not necessarily available in the grid of epochs $\set{T_k}_k$. In \cite{foreman2016}, a similar issue is solved by using the nearest neighbors in the epochs grid. Instead, to allow the efficient matrix computation of \autoref{eq:result}, we interpolate the likelihood grid from $\set{T_i}_i$ to a common grid of transit phases $\set{\phi_i}_i$, leading to the periodic search log-likelihood
$$\ln\mathcal{P}(P) = \set{\ln p(\bm{f} \vert P, \phi_i, D_j)}_{i,j}$$
shown for few periods in \autoref{fig:periodic_search}. In the latter equation, $\Delta_{i,j}$ is omitted since being interpolated from the linear search using $\phi_i$, $D_j$ and $T_0 = 0$.

\subsection{The transit-search periodogram}\label{periodogram}

Using \autoref{eq:result}, we can now compute $\ln\mathcal{P}$ for a range of periods and phases, and build a transit search periodogram using \autoref{eq:periodogram}. This has two disadvantages: First, each likelihood $p(\bm{f} \vert T_0, D, \Delta)$ estimated during the linear search is computed using N measurements. Hence, combining transits in the periodic search, through $\Delta_k$, $\sigma_k$ and the product of $K$ likelihoods $\set{\mathcal{L}_k}_k$ (see \autoref{eq:result}), artificially leads to a likelihood involving up to $N\times K$ measurements. For this reason, one has to normalize each likelihood $\ln\mathcal{P}$, by keeping track of the number of points used to compute each of them, which differs from one phase to another. Second, the maximum value of the likelihood $\ln\mathcal{P}$ is relative to a given dataset, so that a more intuitive and absolute metric must be used to relate to the transit signal detection (such as the Signal Detection Efficiency in \citealt{bls}). This motivates a final step to produce an interpretable transit search periodogram $\mathcal{Q}$. 
\begin{figure}[H]
    \begin{centering}
        \includegraphics[width=\linewidth]{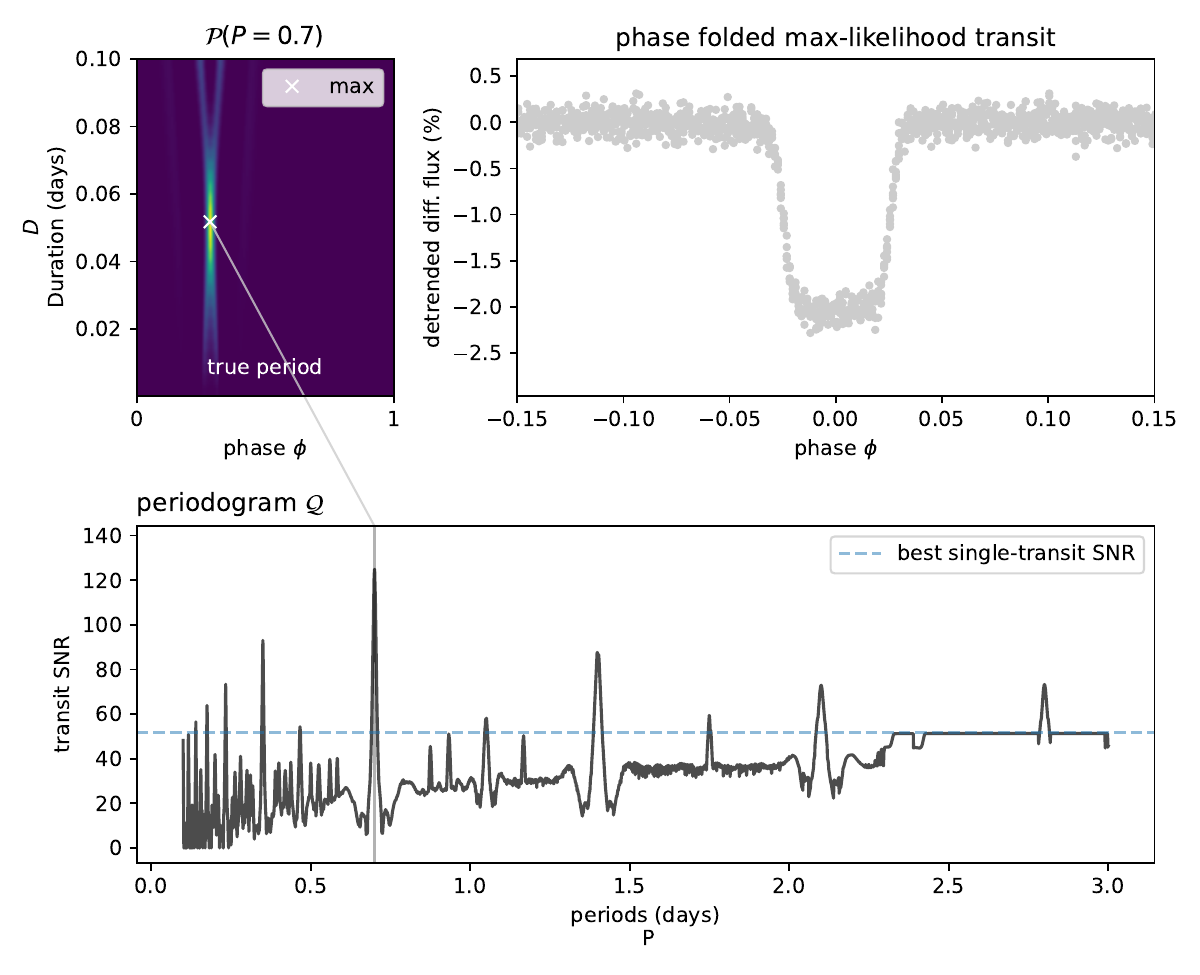}
        \caption{For each period $P$, the joint likelihood $\mathcal{P}(P)$ is computed using \autoref{eq:result}, and the value of the maximum likelihood transit SNR retained as $\mathcal{Q}(P)$. \codelink{workflows/principle/principle.ipynb}}
        \label{fig:periodogram}
    \end{centering}
\end{figure}
For any period P, instead of taking $ \mathcal{Q}(P)$ as the maximum value of $\ln\mathcal{P}$, we compute the maximum likelihood parameters
\begin{equation}\label{eq:phi0}
    (\phi_0 ,D) = \argmax_{\phi_i, D_j} \set{\ln p(\bm{f} \vert P, \phi_i, D_j)}_{i, j}
\end{equation}
and define $\mathcal{Q}(P)$ as being equal to the SNR of the transit of period $P$, epoch $T_0 = \phi_0 \times P$, duration $D$ and depth $\Delta$, i.e.
\begin{equation*}
    \mathcal{Q}(P) = \frac{\Delta}{\sigma},
\end{equation*}
where $\Delta$ and $\sigma$ are obtained using \autoref{eq:result} with the last column of $X$ containing a periodic transit signal of period $P$, epoch $T_0$, duration $D$ and depth $1$. This process and the resulting periodogram $\mathcal{Q}$ are shown in \autoref{fig:periodogram}.\\\\
Hence, periodic transit of period $P$ with the maximum SNR, i.e. maximizing $\mathcal{Q}$, is adopted as the best candidate, basing the confidence in this signal through its SNR. The parameters of this transit are the period $P$, epoch $T_0 = \phi_0 P$, duration $D$ (\autoref{eq:phi0}), and depth $\Delta$ with error $\sigma$ (given by \autoref{eq:result}).\\\\
\subsection{An open-source python package}\label{package}
The methods presented in this paper are made available through the \nuancecode{} open-source Python package hosted at \href{https://github.com/lgrcia/nuance}{https://github.com/lgrcia/nuance}, released on the Python Package Index\footlink{https://pypi.org/project/nuance/} and with documentation and tutorials hosted at \href{https://nuance.readthedocs.io}{https://nuance.readthedocs.io}. All following mentions of \nuancecode{} refer to its \review{version 0.6.0 \citep{zenodo}.}{} \\\\
To instantiate a search, a user can start by creating a \textsf{Nuance} object with
\begin{lstlisting}[language=Python]
from nuance import Nuance

nu = Nuance(time, flux, gp=gp, X=X)
\end{lstlisting}
where \textsf{gp} is a \textsf{tinygp} GP instance and \textsf{X} the design matrix of the linear model. \textsf{nuance} exploits the use of \textsf{tinygp}\footnote{\href{https://github.com/dfm/tinygp}{https://github.com/dfm/tinygp}}, a Python package powered by \textsf{JAX}\footnote{\href{https://github.com/google/jax}{https://github.com/google/jax}}, allowing for custom kernels to be built and highly tractable computations. We can then define a set of epochs \textsf{t0s} and durations \textsf{Ds} and run the linear search with
\begin{lstlisting}[language=Python,linewidth=\linewidth]
import numpy as np

t0s = time.copy()
# a range of 10 durations
Ds = np.linspace(0.01, 0.2, 10)
nu.linear_search(t0s, Ds)
\end{lstlisting}
Finally, the periodic search is run with
\begin{lstlisting}[language=Python]
# range of periods
periods = np.linspace(0.1, 5, 2000)
search = nu.periodic_search(periods)
\end{lstlisting}
From this \texttt{search} object, the best transiting candidate parameters can be computed (\textsf{search.best}), or the $\mathcal{Q}$ periodogram retrieved (\textsf{search.Q\_snr}), together with valuable information about the transit search. The \textsf{Nuance} object also provides methods to perform transit search on light curves from multi-planetary hosts, the advantage of \nuancemethod{} being that the linear search only needs to be performed once and reused for the search of several transiting candidates (see \autoref{toi540}). An extensive and maintained online documentation is provided at \href{https://nuance.readthedocs.io}{\texttt{nuance.readthedocs.io}}.

\subsection{Comparison with BLS}\label{control}

\begin{figure}[H]
    \begin{centering}
        \includegraphics[width=\linewidth]{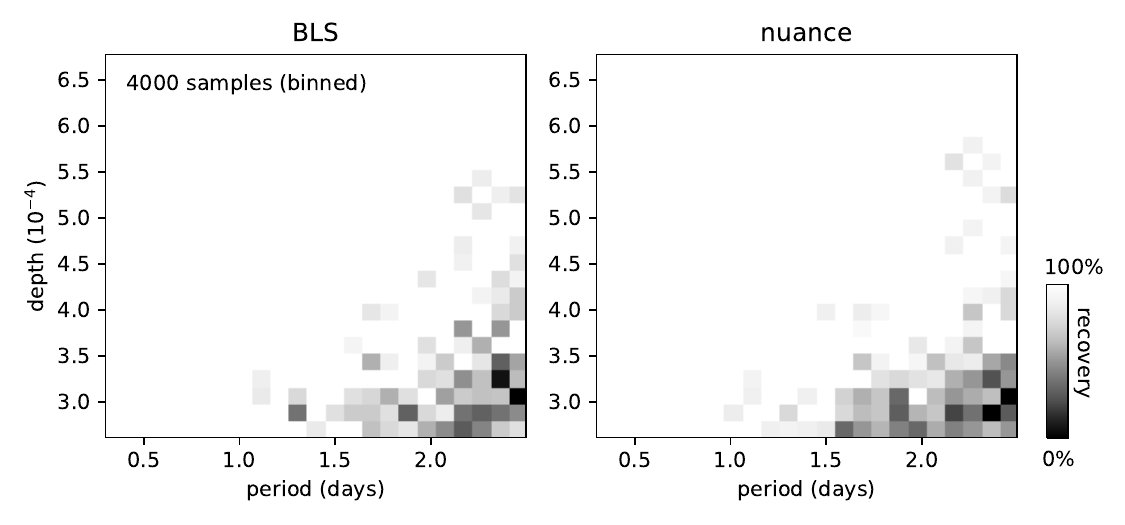}
        \caption{Binned statistics of the injection-recovery of 4000 transit signals in a flat light curve with only white noise using \textsf{BLS} and \nuancecode{}. The color scale indicates the recovery of transits in the corresponding (period, depth) parameter space, white for a full recovery and black for no detection. \codelink{workflows/control_test_bls/scripts/plot.py}}
        \label{fig:control}
    \end{centering}
\end{figure}

To start testing \nuancecode{} against existing methods, a simple adimensional normalized light curve is simulated, consisting of pure white noise with a standard deviation of $5\times 10^{-4}$ spanning 6 days with an exposure time of 2 minutes.
From this signal, we produce 4000 light curves, each containing box-shaped transits with periods randomly sampled from 0.3 to 2.5 days, durations of 50 minutes, and depths randomly sampled to lead to transit SNRs ranging from 4 to 30.
% (i.e.\;depths from $2.6 \times 10^{-4}$ to $6.8 \times 10^{-4}$, using the simple model from \cite{protopapas} presented in \autoref{app_proto} with $c=500$). By design, these injected signals have an SNR ranging from 5 to 30 (with $\sigma_r = 0$ in \autoref{eq:snr}). 
For each light curve, transits are searched using two different tools: \textsf{nuance}, using its implementation from the Python package described in the previous section; and BLS, the Box-Least-Squares algorithm from \cite{bls} (using \textsf{astropy}'s \textsf{BoxLeastSquares}\footlink{https://docs.astropy.org/en/stable/api/astropy.timeseries.BoxLeastSquares.html} implementation). \review{We note that the Transit-Least-Squares algorithm \citep{tls}, that includes the effect of limb-darkening in the base transit template, could also be used here. However, as this effect has a negligible impact on transit detection compared to the effect of correlated noise, we choose to make our comparisons with BLS only.}\\\\
For both methods, 3000 trial periods from 0.2 to 2.6 days are searched, with a single trial duration fixed to the unique known duration of 50 minutes. A transit signal is considered detected if the absolute difference between the injected and the recovered period is less than 0.01 day. To ease the detection criteria, orbital periods recovered at half or twice the injected ones (aka \textit{aliases}) are considered as being detected. For this reason, detected transit epochs are not considered (although manually vetted). Results from this \textit{injection-recovery} procedure are shown in \autoref{fig:control}.\\\\
These results demonstrate the qualitative match between the detection capabilities of \textsf{nuance} and \textsf{BLS} on light curves with no correlated noise, where the BLS algorithm should be optimal. Explaining the subtle differences observed between the two methods when only white noise is present is beyond the scope of this paper, and we will assume that any differences observed in the following sections are due to the different treatments of correlated noise.

\newpage
\section{Performance}\label{results}
\autoref{fig:snr_detrend} shows that \textsf{nuance}'s full-fledged modeling capabilities may not always be necessary and may only be beneficial for certain noise characteristics, relative to the searched transit parameters. Here, we evaluate the performance of \textsf{nuance} in the relative parameter space $(\tau, \delta)$ described in \autoref{eq:relative_params} and probe when its specific treatment of correlated noise in the transit search becomes necessary.\\\\
We perform this study by comparing \textsf{nuance} to the approach that involves removing stellar variability from light curves before performing the search on a detrended dataset. The following detrending strategies, each followed by a search with the BLS algorithm, are compared:\\\\
\ul{\texttt{bi-weight+BLS}}: employs an optimal bi-weight filter implemented in the \textsf{wõtan} Python package with an optimal window size of $3\times D$, i.e.\, three times the transit duration (e.g.\;\citealt{wotan} and \citealt{Dransfield2024}).\\\\
\ul{\texttt{GP+BLS}}: employs a GP conditioned on the data (e.g.\;\citealt{Lienhard2020}). The kernel of the GP and its optimization is described on a case-by-case basis.\\\\
\ul{\texttt{Bspline+BLS}}: employs a B-spline\footlink{https://docs.scipy.org/doc/scipy/reference/generated/scipy.interpolate.BSpline.html} for detrending, fitted using the \textsf{scipy.interpolate.splrep} function\footlink{https://docs.scipy.org/doc/scipy/reference/generated/scipy.interpolate.splrep.html} (e.g.\;\citealt{wotan} and \citealt{Canocchi2023}).\\\\
\ul{\texttt{harmonics+BLS}}: employs a linear harmonic detrending, where the light curve is modeled as a Fourier series including four harmonics of the stellar rotation period with coefficients found through ordinary least squares.\\\\
\ul{\texttt{iterative+BLS}} iteratively detrends the light curve with a sinusoidal signal fitted to the data, each time using the dominant period of the residuals found using a lomb-scargle periodogram (5 iterations).\\\\
Like in previous sections, we consider the planet detected if the recovered period is within 0.01 day of the true period or a direct alias (such as two times or half the true period). Again, we ignore the exact match between the injected and recovered transit epochs, although we visually vet that the found epochs are consistent with the ones injected. We consider a transit detectable if its original SNR is greater than 6. Hence we define \textit{true positives} as detectable transits recovered with the correct period (or an alias) and a measured SNR greater or equal to 6, and \textit{false positives} as undetectable transits recovered with a measured SNR greater or equal to 6.\\\\As we noticed that few methods were still affected by the remaining stellar variability after detrending, the grid of orbital periods being searched for does not contain the stellar rotation period $P_*$ and its aliases. In practice, this is done by removing all orbital periods $P$ from the search grid such that $dP = \frac{P}{P_*}$ is less than 2\% from an integer value, i.e. $\vert  dP - \lceil dP \rceil \vert < 0.02$.\\\\
\subsection{Comparisons on simulated light curves}\label{simu}
Our first comparison dataset consists of 4000 light curves simulated using the model described in \autoref{light_curves_simulations}. We simulate a common periodic transit added to all light curves, of period $P=1.1$ days, epoch $T_0=0.2$ days, duration $D=0.04$ days and depth $\Delta=1\%$. Each light curve consists of a 4 days observation with an exposure time of 2 minutes, leading to $N=2880$ data points with a normal error of $0.1\%$.\\\\
For a given pair of $(\tau, \delta)$, we simulate stellar variability using a GP with an SHO kernel of hyperparameters defined by \autoref{eq:relative_params} computed with respect to the injected transit parameters $D$ and $\Delta$. The same kernel is used for the search with \textsf{nuance} and with the \texttt{GP+BLS} method, an optimal choice on equal footing with the optimal $3\times D$ window size of the bi-weight filter employed in the \wtls{} search.
The 4000 pairs of $(\tau, \delta)$ are drawn from
\begin{equation*}
    \tau \sim \mathcal{U}(0.1, 10), \hspace{0.5cm} \delta \sim \mathcal{U}(0.1, 25) \hspace{0.5cm}\text{and}\hspace{0.5cm} Q \sim \mathcal{U}(10, 100)
\end{equation*}
where $\mathcal{U}(a, b)$ denotes a uniform distribution of lower bound $a$ and upper bound $b$.\\\\
\subsubsection*{Results}
\begin{figure}[H]
    \begin{centering}
        \includegraphics[width=\linewidth]{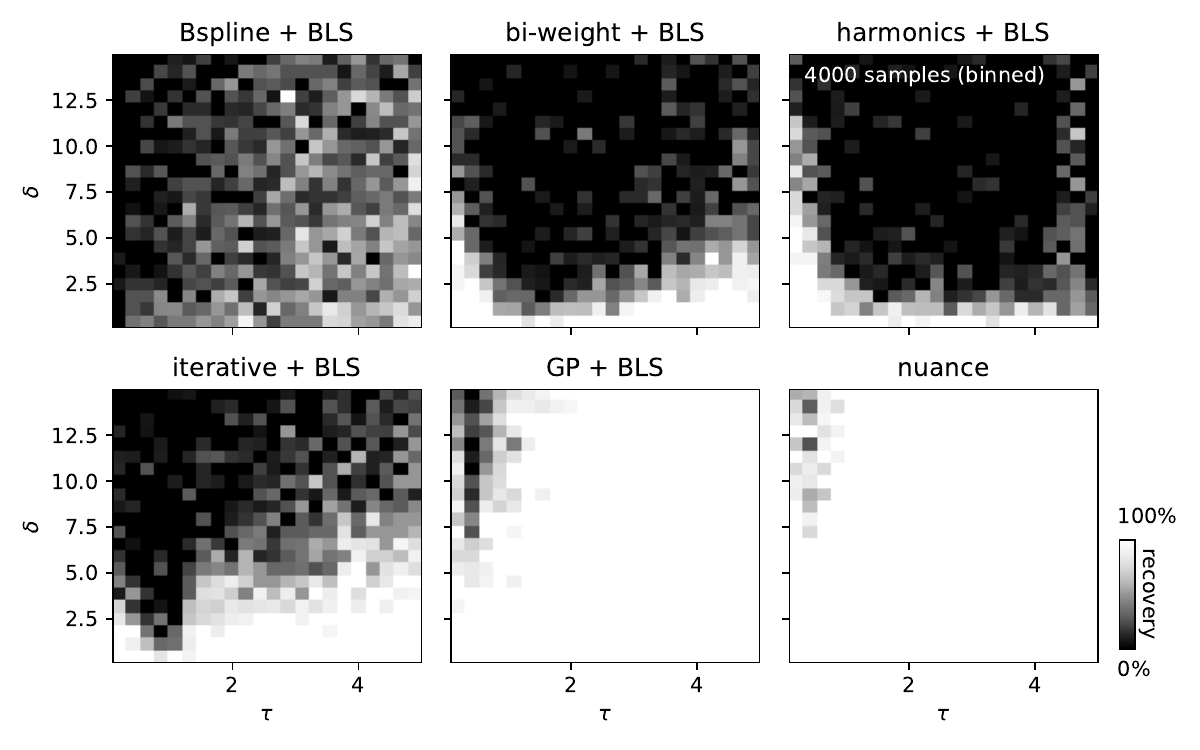}
        \caption{Rate of true positives for the 4000 simulated light curves. The color scale represents the fraction of true positives, white if all injected transits are recovered in a given portion of the $(\tau, \delta)$ parameter space, black if none are recovered. \codelink{workflows/synthetic-injection-recovery/scripts/plot.py}}
        \label{fig:simu}
    \end{centering}
\end{figure}

The results of this injection-recovery procedure are shown in \autoref{fig:simu} and highlight particularly well the benefit of \textsf{nuance} against most other methods. Except for the \texttt{GP+BLS} approach, \nuancecode{} leads to a much higher rate of true positives for transits with relatively small depths compared to the stellar variability amplitude (i.e.\;$\delta>2$), and a duration comparable to the stellar variability period (i.e.\;$\tau<2$). On the other hand, the \texttt{GP+BLS} strategy seems almost as performant as \nuancecode, recovering most of the injected transits and only lacking detection in a relatively comparable portion of the $(\tau, \delta)$ parameter space. We note that these empirical statements only concern simulated light curves with a given amount of white noise, and may vary depending on the length of the observing window or the number of transits. For this reason, quantifying for which values of $(\tau, \delta)$ \textsf{nuance} outperform these methods would only apply to this specific but representative example. However, we verify that our conclusions remain qualitatively valid for various simulation setups.

\noindent This injection-recovery is done in a particularly optimal setup, on simulated light curves that are not all physically realistic and using an optimal GP kernel, hence demonstrating the performance of \textsf{nuance} only on a purely \review{synthetic}{} basis. \review{Hence, we strongly emphasize that these tests do not reflect well the performance of each method on real datasets}. In the next section, we perform transits injection-recovery on real space-based light curves.\\\\
\subsection{Comparisons on rapidly-rotating M dwarfs TESS light curves}\label{real}
In order to assess the performance of \nuance{} on real datasets, we inject and recover transits into light curves from the Transiting Exoplanet Survey Satellite (TESS, \citealt{tess}). We focus this proof of concept \review{on the light curves of 438 M-dwarfs found to have detectable rotation signals with periods lower than one day}{} \citep{Ramsay2020}, which lead to a parameter space justifying the use of \nuance{}. For each \review{of the 438}{} targets, transits are injected and recovered in the TESS 2 min cadence SPOC Simple Aperture Photometry and Pre-search Data Conditioning light curves (PDCSAP, \citealt{spoc}) of a single sector (the first being observed for each target) spanning on average 10 days. To our knowledge, none of these 438 targets have been searched for planetary transits before. However, we note that the presence of existing transit signals in these light curves before the injection of simulated ones is possible, but will not affect the relative  comparison of one method to another. \review{As described in the following sections, our experiment considers a total of $438 \times 100$ light curves, where transit signals are injected and recovered.}\\\\
\subsubsection*{Light curve cleaning and transits injection}
As some of the techniques compared to \nuance{} can be affected by gaps in the data, we only use continuous measurements from half a TESS sector. We assume that all methods (including \nuance{}) are based on an incomplete model of the data that does not account for stellar flares. For this reason, the light curve of each target is cleaned using an iterative sigma clipping approach.
\begin{figure}[H]
    \centering
    \includegraphics[width=\linewidth]{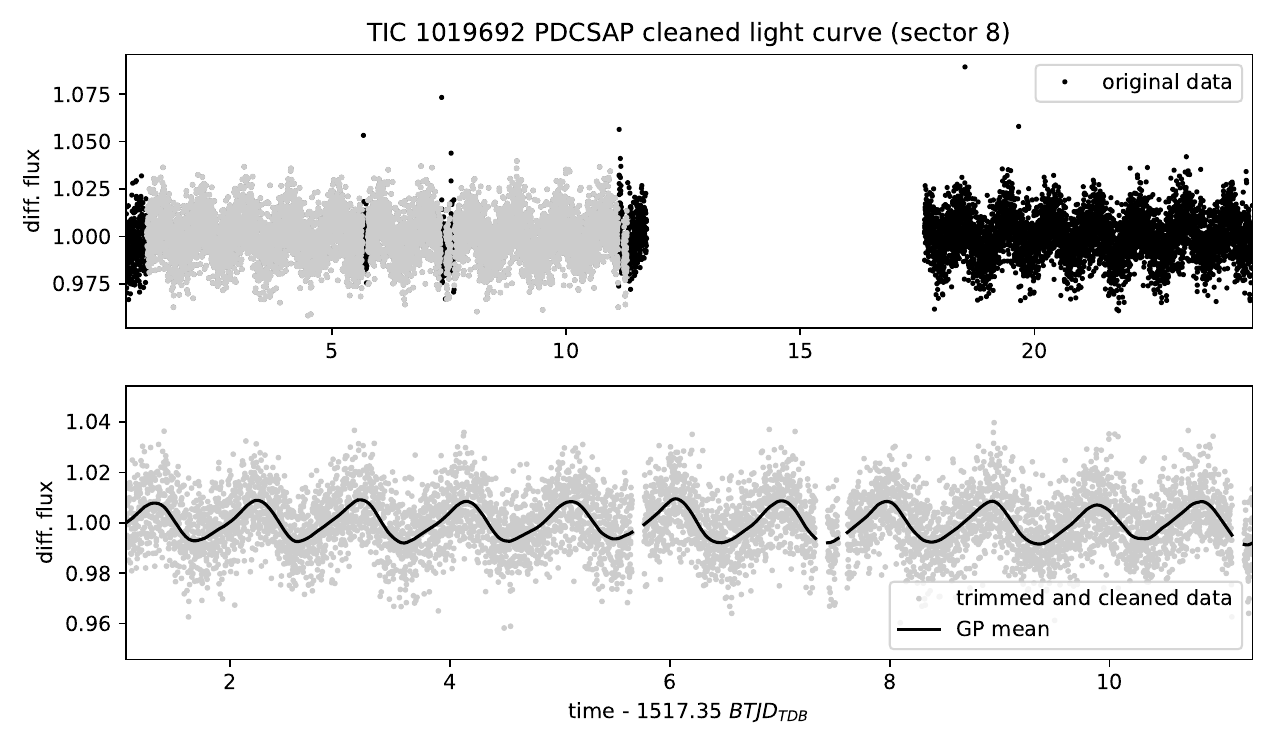}
    \caption{Trimmed and cleaned single-sector light curve of the target TIC 1019692. The top plot shows how much of the data is truncated and sigma clipped, resulting in a quasi-continuous light curve shown in the bottom plot. On this bottom plot, the black line corresponds to the mean of the GP model (with hyperparameters optimized here on the cleaned light curve). \codelink{workflows/tess_injection_recovery/scripts/plot_lc.py}}
    \label{fig:cleaned}
\end{figure}
For each iteration, points 3 times above the standard deviation of the full light curve (previously subtracted by its median) are identified. Then, the 30 adjacent points on each side of the found outliers are masked. This way, large flare signals are masked, using a total of 3 iterations. At each iteration, the GP kernel hyperparameters are re-optimized. As PDCSAP light curves often start with a ramp-like signal, the first 300 points (as well as the last 300 points) of each continuous observation are masked. Finally, each light curve is normalized by its median value. Such a cleaned light curve is shown in \autoref{fig:cleaned}. We note that the gaps left after sigma clipping may be problematic for some of the detrending techniques (such as \texttt{bspline+BLS}). However, adopting this flare cleaning step and analyzing light curves with few small gaps is a practice commonly found in the literature.\\\\
For each \review{of the 438 light curves considered}, transits of planets with 10 different orbital periods combined with 10 planetary radii are individually considered, for a total of 100 periodic transits injection-recovery per target. Orbital periods $P$ are sampled on a regular grid between 0.4 and 5 days, and planetary radii $R_p$ are sampled on a regular grid designed to yield a minimum transit SNR of 2 and a maximum of 30. Using \autoref{eq:snr} with $\sigma_r = 0$, the planetary radius leading to a transit with a desired SNR $s$ is given by
\begin{equation*}
    R_p = R_{\star}\,n^{-\frac{1}{4}} \sqrt{\sigma s}\ %\sqrt[\leftroot{0}\uproot{2}4]{n}}. 
\end{equation*}
where $\sigma$ is equal to the mean uncertainty estimated by the SPOC pipeline, $R_\star$ is the radius of the star reported by \cite{Ramsay2020} and $n$ is the number of points in transit computed using a transit duration assuming a circular orbit. \review{In total, $438 \times 100$ light curves are produced and searched for transits, corresponding to planet radii ranging from $0.46 R_\oplus$ to $12.29 R_\oplus$ and a median radius of $2.40 R_\oplus$.}\\\\
\subsubsection*{Stellar variability kernel}\label{rotation_kernel}
In the \texttt{GP+BLS} and \nuancecode{} methods, we model light curves using a GP with a mixture of two SHO kernels of period $P_\star$ and $P_\star/2$ where $P_\star$ is the rotation period of the star. This model is representative of a wide range of stochastic variability in stellar time series\footlink{https://celerite2.readthedocs.io/en/latest/api/python} \citep[e.g.][]{David2019, Gillen2020}. In order to account for additional correlated noises, we complement this kernel with a short and a long-timescale exponential term, so that the full kernel can be expressed as
\begin{equation*}
    k = k_1 + k_2 + k_3 + k_4
\end{equation*}
with
\begin{itemize}
    \item $k_1$ a SHO kernel with hyperparameters \begin{equation*}
        Q_1 = 1/2 + Q_0 + \delta Q\,, \hspace{0.5cm}
        \omega_1 = \frac{4\,\pi\,Q_1}{P\,\sqrt{4\,Q_1^2 - 1}} \hspace{0.5cm} \text{and} \hspace{0.5cm}
        S_1 = \frac{\sigma^2}{(1 + f)\,\omega_1\,Q_1}.
    \end{equation*}
    \item $k_2$ a SHO kernel with hyperparameters \begin{equation*}\begin{gathered}
        Q_2 = 1/2 + Q_0\,, \hspace{0.5cm}
        \omega_2 = 2 \omega_1 \hspace{0.5cm} \text{and} \hspace{0.5cm}
        S_2 = \frac{f\,\sigma^2}{(1 + f)\,\omega_2\,Q_2},
    \end{gathered}\end{equation*}
\end{itemize}
where $Q_0$ is the quality factor for the secondary oscillation, $\delta Q$ is the difference between the quality factors of the first and the second modes, $f$ is the fractional amplitude of the secondary mode compared to the primary and $\sigma$ is the standard deviation of the process. The kernels $k_3$ and $k_4$ are expressed as 
\begin{equation*}
        k(t, t')=\sigma^2\,\exp\left(-\frac{\vert t - t' \vert}{\ell}\right),
\end{equation*}
with $\ell$ and $\sigma$ the scale and standard deviation of the process, respectively. These are meant to model short and long-timescale nonperiodic correlated noise. In total, the rotation kernel $k$ has 8 hyperparameters.\\\\
The hyperparameters of this kernel are optimized on trimmed and cleaned light curves containing the injected transits, using the \textsf{scipy.optimize.minimize} wrapper provided by the \textsf{jaxopt} Python package\footlink{https://jaxopt.github.io/stable/_autosummary/jaxopt.ScipyMinimize.html}, and taking advantage of the \textsf{JAX} implementation of \textsf{tinygp} and its quasiseparable kernels\footlink{https://tinygp.readthedocs.io/en/latest/api/kernels.quasisep.html} \citep{celerite}. As correlated noise is expected to affect the light curve uncertainty estimates performed by SPOC, the diagonal of the full covariance matrix of the data (i.e.\,their uncertainty, assuming homoscedasticity) is held free, increasing the number of optimized parameters to 9. The optimization is performed using the \textsf{BFGS} algorithm \citep{Fletcher1987}, minimizing the negative log-likelihood of the data as expressed in \autoref{eq:linear_search_ll} (without transit), i.e.\,accounting for a linear systematic model of the data in addition to stellar variability. For simplicity, and to adopt a uniform treatment for all target light curves, a design matrix $\bm{X}$ with a single constant column is adopted, such that the systematic model only consists of a single parameter corresponding to the mean value of the differential flux (expected to be close to 1) solved linearly. Our motivations for choosing this very simplistic baseline, despite the capability of \nuancemethod{} to account for more complex linear models, is discussed in \autoref{systematics}.\\\\
\subsubsection*{Search parameters and transit detection criteria}
For all techniques, we only search for transits with a duration fixed to the known duration of the injected transits. This is mainly done for computational efficiency and allows for a narrower comparison between \nuancemethod{} and all BLS-based techniques. Finally, the search is done over 4000 trial periods linearly sampled from 0.3 to 6 days. A realistic transit search on a wider parameter space (e.g. multiple trial durations) is discussed in the next section.
\newpage
\subsubsection*{Results}
\begin{figure}[H]
    \begin{centering}
        \includegraphics[width=\linewidth]{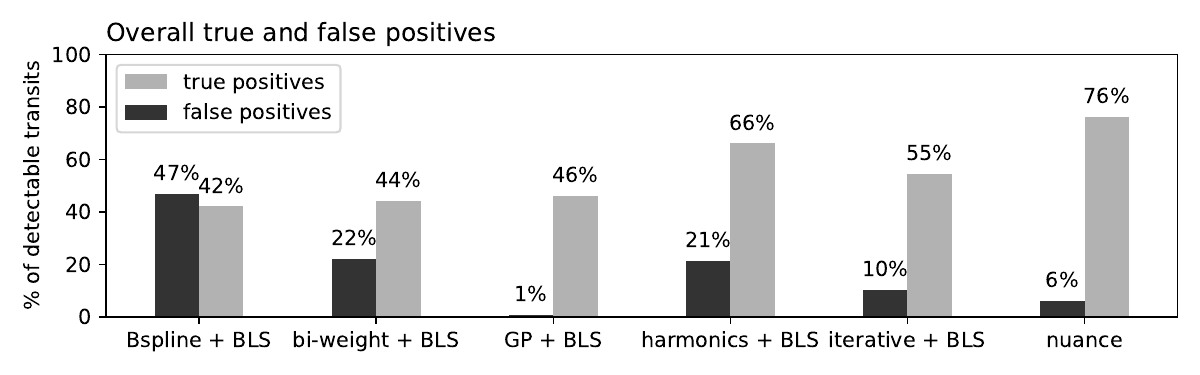}
        \caption{Rate of true and false positives for all methods as a fraction of detectable transit signals. \codelink{workflows/tess_injection_recovery/scripts/plot_tpfp.py}}
        \label{fig:allsearch}
    \end{centering}
\end{figure}

\begin{figure}[H]
    \begin{centering}
        \includegraphics[width=\linewidth]{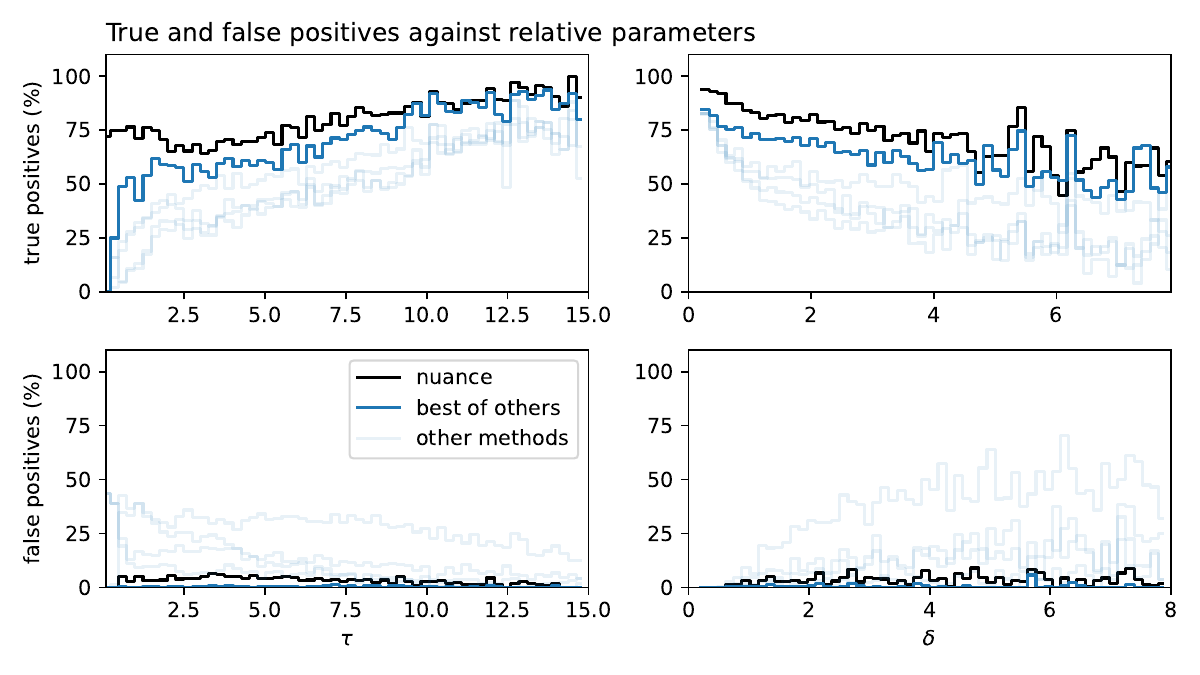}
        \caption{Rate of true and false positives of \nuancecode{} compared to other methods, as a function of the relative parameters $\tau$ and $\delta$. The solid blue line corresponds to the maximum of true positives among all methods on the top panel (dominated \review{by the \texttt{harmonics+BLS} method}), and the minimum of false positives on the bottom panel (waned by the \texttt{GP+BLS} method, all other methods being above \nuancecode{}). \codelink{workflows/tess_injection_recovery/scripts/plot_tpfp.py}}
        \label{fig:allsearch_relative}
    \end{centering}
\end{figure}

\begin{figure}[H]
    \begin{centering}
        \includegraphics[width=\linewidth]{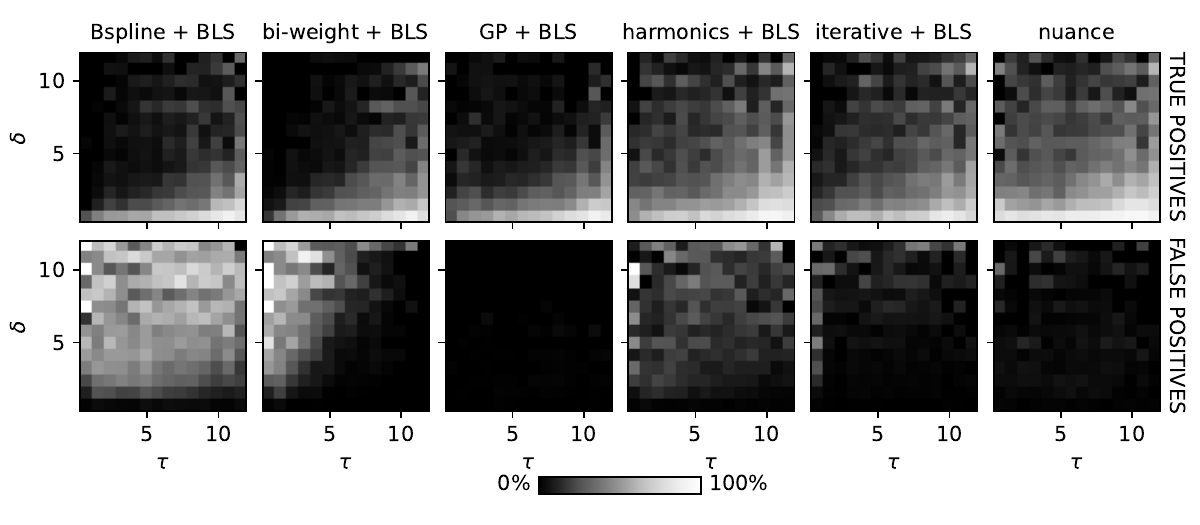}
        \caption{Rate of true and false positives binned over the full parameter space $(\tau, \delta)$ for all methods and all targets. \codelink{workflows/tess_injection_recovery/scripts/plot_tpfp.py}}
        \label{fig:allsearchim}
    \end{centering}
\end{figure}

\begin{figure}[H]
    \begin{centering}
        \includegraphics[width=0.6\linewidth]{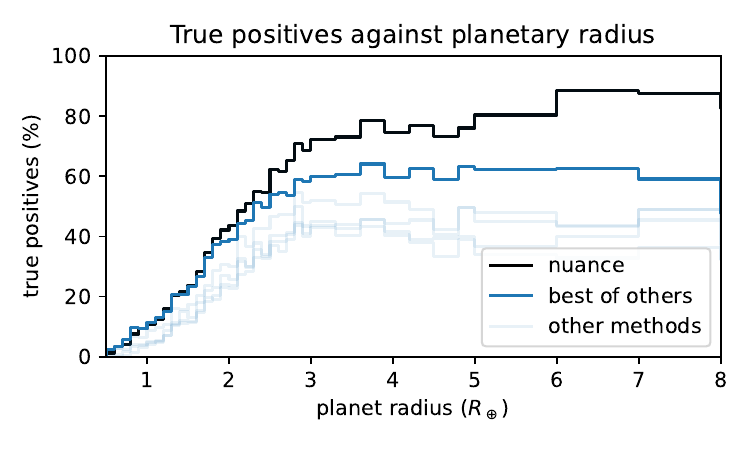}
        \caption{\review{Rate of true and false positives of \nuancecode{} compared to other methods, as a function of the injected planet's radius. The solid blue line corresponds to the maximum of true positives among all other methods than \nuance{} (dominated by the \texttt{harmonics+BLS} method).}{} \codelink{workflows/tess_injection_recovery/radii.ipynb}}
        \label{fig:tpfp_radius}
    \end{centering}
\end{figure}

An example of the transit injection-recovery and its result is shown in \autoref{1019692} for the target TIC 1019692, \review{a representative example of the kind of stellar variability encountered in our dataset}. \autoref{fig:allsearch} shows the global results of the injection-recovery for all targets, while \autoref{fig:allsearch_relative} shows the true and false positives plotted against the relative parameters $\tau$ and $\delta$ (defined in \autoref{eq:relative_params}). These figures are a synthesis of \autoref{fig:allsearchim}, which shows the rate of true and false positives binned in the full parameter space $(\tau, \delta)$.
Compared to other techniques, we find that \nuancecode{} leads to the highest number of true positives, with a successful detection of 76\% of the 7008 detectable transits injected (\autoref{fig:allsearch}). While the performances of other methods strongly depend on the characteristics of the variability, \textbf{\nuancecode{} is the best technique in 93\% of cases, leading to both the highest number of true positives and the lowest number of false positives} (\autoref{fig:allsearch}).\\\\
From \autoref{fig:allsearch}, we note that the number of true positives of the \texttt{GP+BLS} method is much worse than what could be anticipated from the results presented in \autoref{fig:simu}, where the GP kernel was fully optimal (given it was also used to simulate the data). In that case, our kernel might not be optimal, whether because of its form or because of the values of its hyperparameters. In any case, the fact that the same kernel performs significantly better when used with \nuancecode{} shows that \textbf{our method is less sensitive to the choice of kernel and its optimization compared to detrending with a GP}.\\\\
\review{Finally, in \autoref{fig:tpfp_radius}, we observe a dependence between the rate of true positives and the injected planets radii. Overall \texttt{nuance} leads to significantly higher true positives in all radii bins compared to other methods, except for \texttt{BLS+harmonics} (which dominates other methods), which shows a similar yield for planets with radii less than 2$R_\oplus$. Hence, we note that the advantage of \texttt{nuance} is particularly significant for planets larger than 2$R_\oplus$. This can be explained by the fact that, for a transit signal to be comparable to a high-amplitude stellar variability signal, i.e. when \texttt{nuance} shows an advantage, a larger transit depth, hence a larger planetary radius, is required. Although this particular result highly depends on the variability present in the studied dataset, we conclude that \textbf{the performance of \nuance{} is not sensitive to the searched planets' radii.}}\\\\
\review{When analyzing the light curves of the 438 targets from \citealt{Ramsay2020}, our search solely focused on the injected transit signals. Although \nuancecode{} seems particularly well suited to search for real transit signals in this dataset, we did not conduct such study. To be done properly, this task would simply require a search on a wider and finer trial epochs and durations grid, using all available TESS sectors; a project that we highly encourage.}{}

\subsection{Comparison for a multi-sector TESS candidate: TOI-540}\label{toi540}
\begin{figure}[H]
    \begin{centering}
        \includegraphics[width=\linewidth]{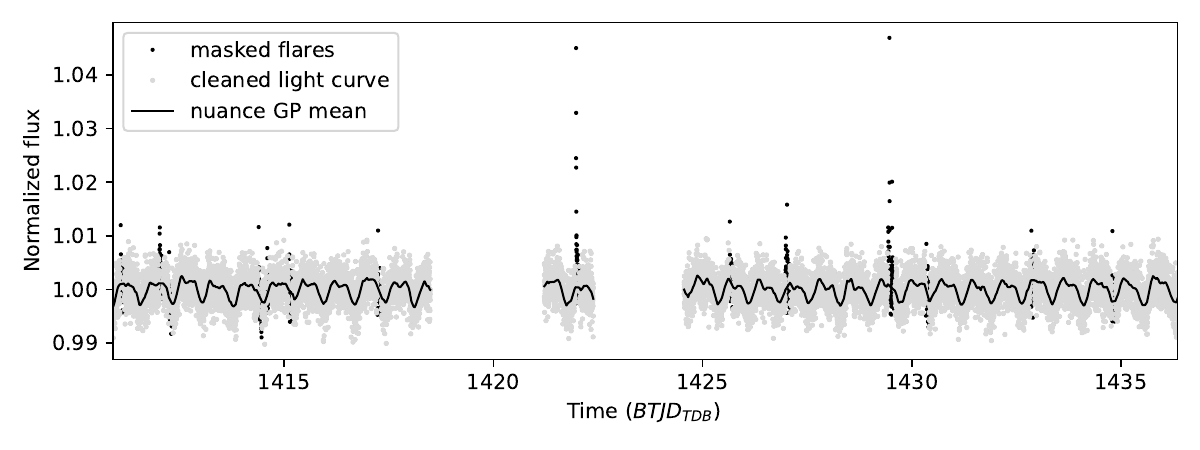}
        \caption{Sector 4 light curve of TOI 540. The cleaned signal (gray points) has been masked for flares (black points), and the black line corresponds to the mean of the GP model. \codelink{workflows/comparison_toi/scripts/plot_light_curve.py}}
        \label{fig:toi540_clean}
    \end{centering}
\end{figure}
In order to further validate nuance on a realistic dataset, we focus this section on the multi-sector TESS light curves of TOI-540, and the search for its Earth-like companion TOI-540 b \citep{TOI540}. \review{As we will see, \nuancemethod{} does not have significantly superior performance for this candidate. However, its light curves feature a very typical stellar variability signal (see \autoref{fig:toi540_clean}), that would commonly be detrended before searching for planetary transits}.\\\\
We downloaded the 2 minute cadence SPOC PDCSAP light curves of TOI-540 observed in 5 sectors (4, 5, 6, 31 and 32). Like in the previous section, we use a Lomb-Scargle periodogram and identify the 0.72 day rotation period of the star, which we use as an initial value to optimize the kernel described in \autoref{rotation_kernel} on each sector independently. Here again, we employ an upper-sigma-clipping to mask flares out of the data. The resulting light curve for sector 4 and its mean model are shown in \autoref{fig:toi540_clean}.\\\\
For each sector, we perform the linear search of nuance on the cleaned light curve, using the original times as the trial epochs and 10 trial transit durations linearly sampled from 15 minutes to 1.5 hours. We then perform the periodic search on all sectors combined, using a concatenation of all linear searches. By adopting this by-sector GP modeling of the light-curve, the linear search of \nuance{} scales linearly with the number of sectors being processed.
\begin{figure}[H]
    \begin{centering}
        \includegraphics[width=\linewidth]{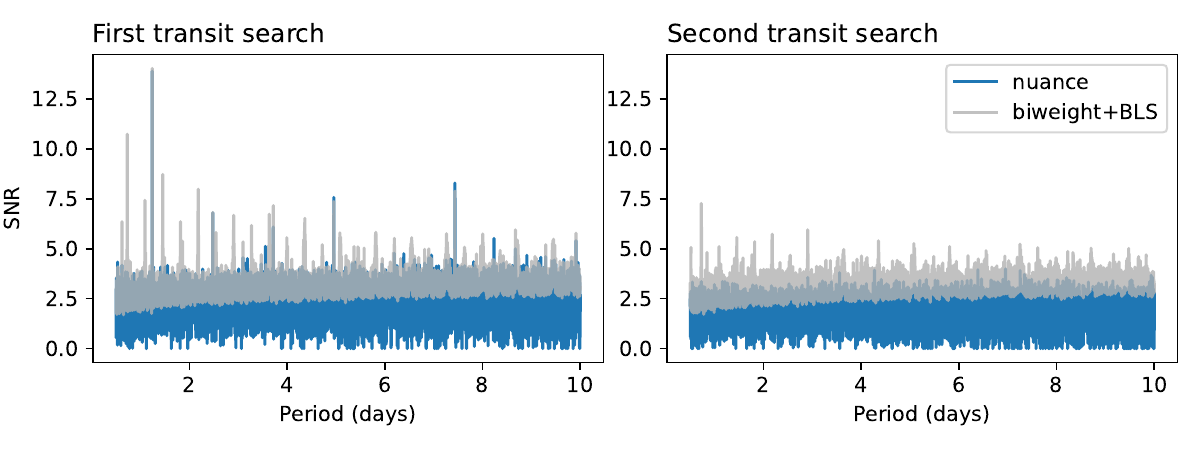} 
        \caption{Transit search SNR periodograms of TOI-540 using \texttt{bi-weight+BLS} and \nuance{}. After a first periodic search (left panel), the epochs corresponding to the maximum-SNR transit are masked before the second search is performed (right panel). \codelink{workflows/comparison_toi/scripts/plot_periodograms.py}}
        \label{fig:toi540_periodograms}
    \end{centering}
\end{figure}
This approach is adopted for efficiency but also to encapsulate the changing properties of stellar variability from one sector to another, often separated by year-long gaps. The periodic search is done on 20 000 trial periods ranging from 0.5 to 10 days\footnote{We acknowledge that the grid of trial periods and durations used in this study is nonoptimal for a real transit search. However, this simple choice is sufficient for our periodogram comparison against BLS. Optimal grids of parameters for realistic transit searches are discussed in \cite{tls}.}. This search, using \nuancecode{}, is compared to the more traditional approach that consists of detrending each sector with a bi-weight filter and then searching for transits with the BLS algorithm (denoted \texttt{bi-weight+BLS} and described in \autoref{real}) on all sectors combined. Since we do not know the transit duration a-priori, we perform the detrending and BLS search using 15 filtering window sizes sampled from 30 minutes to 5 hours, and retain the search that leads to the highest transit SNR peak in the periodogram (as done e.g.\;in the SHERLOCK transit search pipeline described in \citealt{Pozuelos2020}). The results of this comparison are shown in \autoref{fig:toi540_periodograms}.\\\\
After a first periodic search, trial epochs in windows of widths $2\times D$ centered on the detected periodic transits are masked. In practice, this is done by masking the linear search products $\set{\ln\mathcal{L}_{i,j}}_{i, j}$, $\set{\Delta_{i,j}}_{i, j}$ and $\set{\sigma_{i,j}}_{i, j}$ (defined in \autoref{linear_search}).\\\\
As seen in \autoref{fig:toi540_periodograms}, the SNR periodogram using \texttt{bi-weight+BLS} and \texttt{nuance} are very similar, with the known transiting exoplanet TOI-540 b detected with an orbital period P = 1.24 days. This is well expected as the relative parameter $\tau$ equals 13 for TOI-540 b\footnote{Using \autoref{eq:relative_params} with the stellar rotation $P=0.72$ days and the known transit duration $D=29.5$ minutes of TOI-540 b \citep{TOI540}}, which lies outside the range where \nuancemethod{} is expected to be beneficial (see \autoref{fig:snr_detrend}). \review{Nonetheless}, the \nuance{} periodogram of TOI 540 features less spurious SNR peaks, largely due to the penalty naturally occurring when single transits with different depths are periodically combined. In the second search (right panel) of \autoref{fig:toi540_periodograms}, we also notice a higher number of peaks that would lead to false positive detections of transits in the \texttt{bi-weight+BLS} case. The proper treatment of correlated noise in \nuance{}, as observed in \autoref{real}, makes these peaks nonsignificant, avoiding a large number of false detections.\\\\
We note that finding a known TESS candidate that displays characteristics for which \nuance{} is expected to be significantly beneficial proved to be very challenging during the writing of this paper, as transits with such characteristics are expected to be missed by current state-of-the-art techniques (see, e.g., \autoref{fig:simu}). Nonetheless, we verify that \nuancemethod{} is capable of finding a large number of already known transiting exoplanet candidates, in light curves featuring various forms of correlated noises, at least as efficiently as with commonly used techniques. We reserve the search for new transit signals for a follow-up publication.

\newpage
\section{Discussion}\label{discussion}

In the previous section, we demonstrated the capability of \nuance{} to search for synthetic or known transit signals, in simulated or real datasets. Here, we discuss the caveats of this algorithm, the advantages and limitations of the \nuancecode{} implementation, and future prospects for its extension.

\subsection{Processing time}

\begin{figure}[H]
    \begin{centering}
        \includegraphics[width=\linewidth]{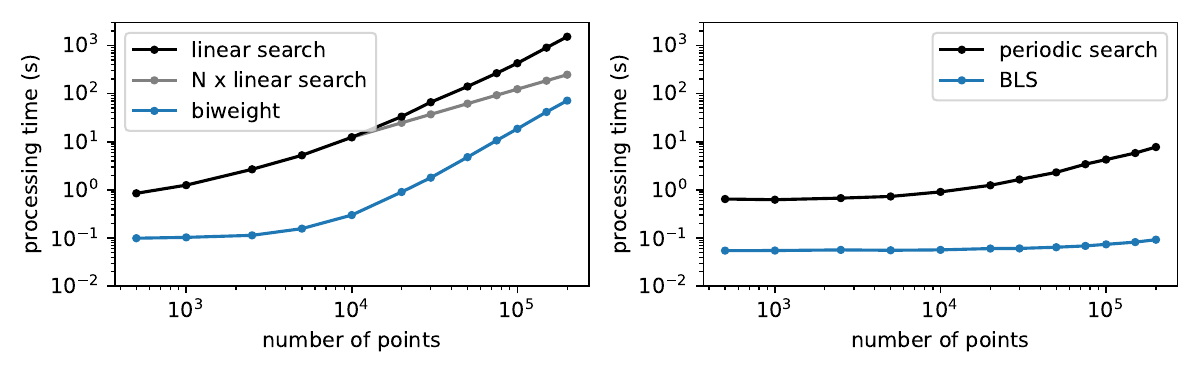}
        \caption{Study of \textsf{nuance} (black line) processing time against \texttt{bi-weight+BLS} (see \autoref{simu}, blue line). The gray curve shows the performance of \textsf{nuance} linear search when applied to chunks of 10 000 points continuous observations, instead of considering these observations all together. This study is performed on a single CPU core of an Apple M2 Max chip. While not being shown, we verify that both the BLS algorithm and \nuancemethod{} processing times scale linearly with the number of trial transit durations and orbital periods. \codelink{workflows/benchmark/scripts/plot.py}}
        \label{fig:blsvsnuance}
    \end{centering}
\end{figure}

\noindent In \autoref{fig:blsvsnuance}, the processing time of \nuancecode{} linear and periodic search are recorded against the number of points in a simulated light curve, assuming a simple un-optimized GP with a squared exponential kernel. These are compared to the processing time of \texttt{bi-weight+BLS} (see \autoref{simu}) separated into the bi-weight filtering step and the BLS search.

As seen in \autoref{fig:blsvsnuance}, most of the computational costs of \nuancecode{} and the \texttt{bi-weight+BLS} method come from the linear search and the bi-weight filtering step. This is not always true and depends on the size of the trial durations and period grids. One advantage of \nuancemethod{} is that the linear search can be performed separately on different continuous observations, and then combined in the periodic search. Hence, if searching for transits in separate observations with approximately similar durations, such as different TESS sectors or different ground-based nightly observations, the computational cost of \nuancecode{} grows linearly with the number of observations (see gray line in \autoref{fig:blsvsnuance}). Nonetheless, considering a variety of other detrending algorithms, \textbf{\nuancecode{} is expected to perform between one and two order of magnitude slower than more conventional techniques associated with BLS}.\\\\ 
Nonetheless, searching for TOI-540 b transits with the parallelized implementation of \nuancecode{} (see \autoref{toi540}) on the 12 cores of an Apple M2 Max chip, took only 5 minutes and 35 seconds, 1 minute 50 seconds for the 5 sectors linear searches (around 22 seconds per TESS sector), and 3 minutes 45 seconds for the combined periodic search. In comparison, the brute force search with \texttt{bi-weight+BLS}, consisting of trying 15 bi-weight windows, took a total of 1 minute 49 seconds (only 7 seconds each).\\\\
% As the periodic search is expected to be improved by lower lever optimizations (using a compiled language), we reserve its benchmark for a future work. Thanks to its \textsf{JAX} implementation, we also note the possibility of running \nuancecode{} on GPUs, which is out of the scope of the present comparison.\\\\
Because of its computational cost, we do not recommend using \nuancecode{} in the general case, but rather when light curves contain correlated noise with specific characteristics. If employing a bi-weight filter for detrending, these characteristics correspond to the ones discussed in \autoref{issues}. But as these strongly depend on the type of detrending technique employed (see \autoref{fig:simu}), we do not provide general guidelines as when \nuancemethod{} should be preferred over a specific detrending technique. To aid users in making an informed choice of algorithm, extensive benchmarks, and guidelines are reserved for future developments and will be progressively shared on \nuancecode{}'s online documentation\footlink{https://nuance.readthedocs.io/en/latest/}.

\subsection{Systematics modeling}\label{systematics}
Throughout the paper, a single-column design matrix $\bm{X}$, corresponding to the mean differential flux (ideally unitary), was employed, hence assuming that the instrumental systematics signals were nonexistent. In practice, \nuancemethod{} has been developed to linearly model systematic signals through more complex design matrices (as in \citealt{foreman2016}), in addition with its capability to model correlated noise while searching for transits. This feature is intentionally unexploited in the comparisons presented in \autoref{results}, as detrending light curves assuming a linear systematics model, such as PLD co-trending vectors \citep{pld}, is highly incomplete if applied on data while ignoring the presence of other astrophysical signals. Comparisons involving more complex design matrices would also be sensitive to the choice of linear components, and would have unwanted repercussions on their results.\\\\
As an illustration, the NEMESIS pipeline \citep{nemesis} starts processing the differential light curves by employing a linear systematics detrending using a least squares fit of the data with a reduced PLD basis, before smoothing the signal from stellar variability using an approach similar to the one employed in the \texttt{bi-weight+BLS} approach (see \autoref{real}), hence detrending the systematics with an incomplete model that does not account for stellar variability. To account for stellar variability while fitting the linear systematics model to the data, a step further would be to use a GP, such as done in the EVEREST \citep{everest2} pipeline. However this would also involve some potential degradation of the transit signals (see e.g.\;\autoref{fig:issue2}), with an hardly distinguishable origin. For these reasons, and to keep our comparisons as targeted as possible, we do not compare commonly used systematics detrending approaches and decided to focus our comparisons solely on stellar variability detrending techniques (although these two aspects often overlap in the literature, e.g. in \citealt{everest1}).\\\\
Although not being demonstrated here, modeling systematics signals while searching for transits on data acquired sparsely is extremely promising for the search of transiting exoplanets, including for ground-based observations that usually suffer from daily interruptions. In this respect, we note the similarity of our linear search (cf\;\autoref{linear_search}) to the one presented in \citealt{Berta2012}, that focused on the detection of single eclipses in the MEarth light curves \citep{Irwin2009}. Similarly, \textbf{\nuancemethod{} would highly benefit the search for transiting exoplanets around M-dwarf type stars}, such as the ones observed by the SPECULOOS survey \citep{speculoos} whose monitoring suffers from both increased red noise (due to atmospheric and instrumental thermal effects discussed e.g.\;in \citealt{Berta2012} and  \citealt{Pedersen2023}) and enhanced stellar variability \citep{Murray2020}. We reserve this promising application for a future study.

\subsection{The GP kernel}
While not being discussed in our study, \textbf{the efficiency of \nuancemethod{} to detect transits in correlated noise might be dependent on the design of its GP kernel}. In the ensemble comparison of \autoref{real}, the goal was to choose a kernel and an optimization strategy suited to most of the studied light curves, leading to few outliers in the results, which were indicative of a badly designed and/or optimized kernel. An alternative, recommended for more realistic blind searches, is to perform model comparison on well selected kernels, and to adapt the optimization strategy to each dataset.\\\\
When using \nuancecode{} on TESS light curves for example, it must be noted that the observed light curve variability might originate from contamination due to several nearby blended stars, so that a physically-interpretable GP kernel representing a single star activity is not necessarily appropriate. On the other hand, a single squared exponential GP kernel might also be sufficient for some applications; an aspect we intend to explore in future applications.\\\\
Something to emphasize is that \nuancemethod{} cannot be used to produce reliable detrended light curves, as an optimal GP is often flexible enough to partially model transits (see \autoref{fig:issue2}). 
In contrast, the idea behind \nuancemethod{} is rather to compute the likelihood of data against a model containing both transits and correlated noise, without ever trying to disentangle both signals. In practice, it means that it can be very hard to actually verify the presence of transits found by \nuancemethod{} visually, so that transits may be detected but remain hidden in correlated noise. This is particularly true for stars displaying very high-frequency photometric pulsations (see the example in \autoref{fig:exoco}).

\subsection{Prospects}
\begin{figure}[H]    
    \includegraphics[width=\linewidth]{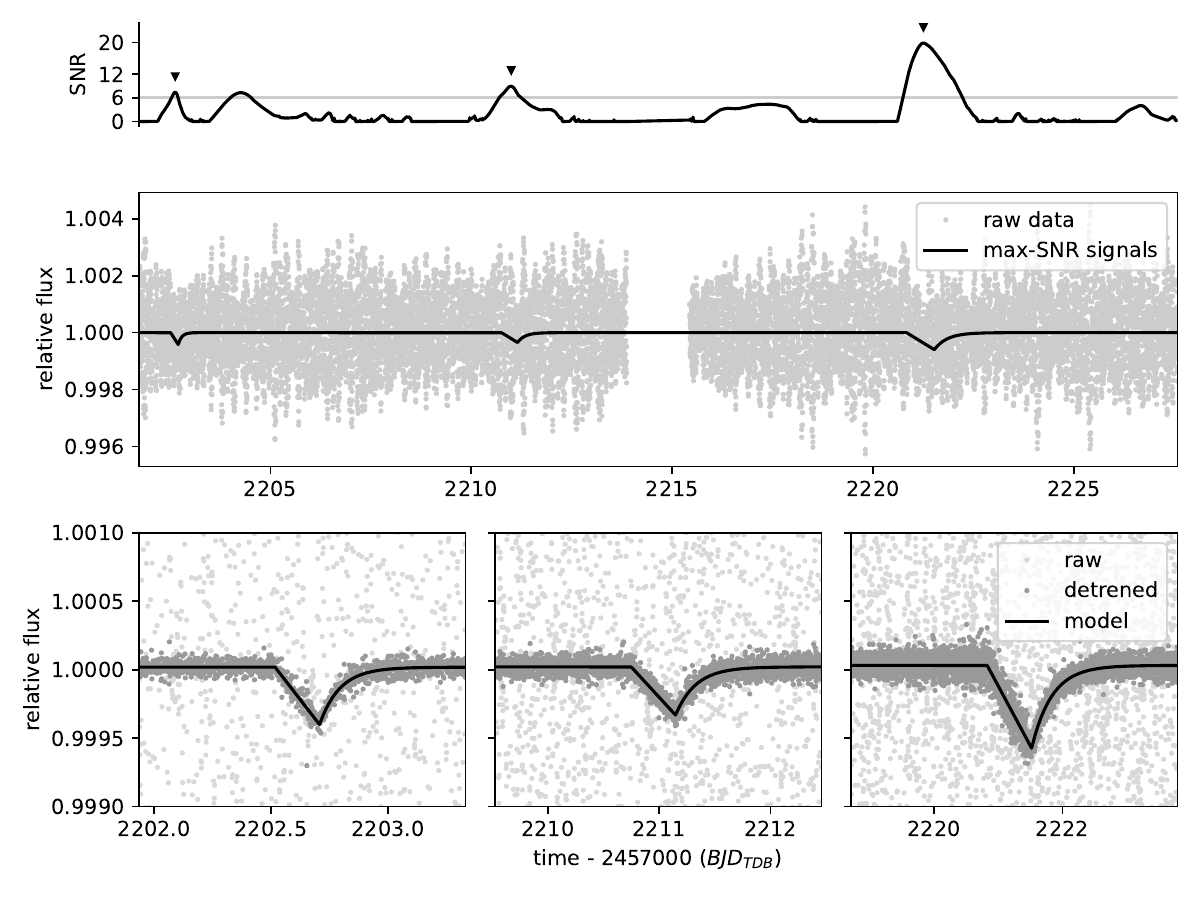}
    \caption{Demonstration of the search of transiting exocomets in the TESS light curves of the star $\beta$ Pictoris (2 minutes PDCSAP data from sector 33). This star is known to display rapid $\delta$ Scuti type photometric variations with a period of about 30 minutes \citep{Lecavelier2022}. The GP kernel and its hyperparameters are chosen and optimized as in \autoref{real}.Here, we simply used the linear search of \nuancecode{} with a different base model, one that mimics the shape of transiting exocomets (with 20 trial durations), to compute the SNR time-series of the signal over one sector. The maximum-SNR events are displayed at the bottom of the figure and match with the ones found by \cite{Lecavelier2022}. \outcodelink{https://nuance.readthedocs.io/en/latest/notebooks/tutorials/exocomet.html}}
    \label{fig:exoco} 
\end{figure}

The present implementation of \nuancemethod{} has the potential to be extended to be used beyond the search of periodic box-shaped transits. Here are ideas of possible use and extensions, from the most straightforward to the most ambitious:
\begin{enumerate}
    \item In order to compare \nuancecode{} to BLS-based methods, we injected and retrieved only box-shaped transits. However, similarly to the Transit-Least-Squares algorithm from \cite{tls}, \textbf{limb-darkened transits can serve as base model} in the linear search and are expected to improve the transit search in the same way TLS provided an improvement over BLS.
    \item The linear search of \nuancemethod{} is a single-event detection algorithm that can be used to search for single transit events, but also \textbf{detect transiting exocomets and flares}, by simply changing the base astrophysical model in the last column of the design matrix $\bf{X}$. While not being tested and benchmarked, \nuancecode{} already integrates this feature. An example of the detection of known transiting exocomets in $\beta$ Pictoris TESS light curves is shown in \autoref{fig:exoco}.
    \item During the periodic search, no prior about the transit duration related to the orbital period of the planet was used. This was done in order to allow the detection of grazing transiting exoplanets that would produce \review{shorter-timescale}{} transits compared to what is expected from a circular orbit with a null impact parameter. However, adding such priors might produce fewer spurious periodogram peaks and be very beneficial for the automatic search of transits in large datasets. Another idea, similar to the one employed in \cite{foreman2016}, is to leverage \textbf{model comparison in order to reject transits that are better described by the GP model alone}. Both ideas come at no cost given our modeling approach.
    % \item \review{The transit-search periodogram was intentionally reduced to a one-dimensional metric, mainly for practical reasons. However, we believe that the profile likelihood (or SNR) over orbital periods \textit{and} transit durations (see \autoref{snrperiod}) holds valuable information that could be exploited to further direct the search, especially when transits with different durations are present.}{}
    \item Finally, the formalism of \nuancemethod{} could be adapted and used to \textbf{search for exoplanets featuring transit time variations (TTVs)}. Indeed, this application only requires a modification of the periodic search, as the maximum likelihood peaks close to linearly predicted transit epochs may be considered. This could be done either with a special nearest-neighbor algorithm or with a convolution of the computed likelihood grid with a Gaussian kernel. To maximize efficiency and interpretability, we would recommend these approaches to be explored analytically, rather than using a data-driven treatment of the linear search products.
\end{enumerate}
% \begin{figure}[H]
%     \begin{centering}
%         \includegraphics[width=\linewidth]{../workflows/principle/figures/principle_snr_period.pdf}
%         \caption{}
%         \label{fig:snrperiod}
%     \end{centering}
% \end{figure}

\newpage
\section{Conclusion}\label{conclusion}
This paper presents \nuancemethod{}, an algorithm designed to detect planetary transits in light curves featuring correlated noise in the form of instrumental signals and stellar variability. In this context, a conventional approach involves detrending a light curve before searching for transits using a Box-Least-Squares algorithm. However, we show that this approach degrades transit signal-to-noise-ratios down to the point of not being detectable. Adopting commonly used detrending strategies, we explore the extent of this degradation on simulated light curves, and its dependence on the photometric stellar variability characteristics, showing the need for a full-fledged transit search method like \nuancemethod{}.\\\\
The effectiveness of \nuancemethod{} is tested using a synthetic dataset and further validated on real TESS light curves of 438 rapidly-rotating M dwarfs. These injection-recovery tests reveal that \nuancemethod{} consistently outperforms commonly used transit search techniques, especially when the timescale of stellar variability is less than twice that of the transit duration. In all cases, \nuancemethod{} not only leads to a higher number of true positive detections but also minimizes false positives, demonstrating its robustness and reliability.\\\\
We make \nuancemethod{} publicly available through the \nuancecode{} open-source Python package, developed with \textsf{JAX} to allow its  use on distributed computing environments and GPU devices. Overall, we acknowledge the limitations of \nuancemethod{} and its increased computational cost compared to more conventional techniques. Hence, \nuancemethod{} should be used as an alternative to more traditional techniques only in the presence of substantial correlated noise. As guidelines for choosing our method over other techniques are highly dependant on the type of detrending algorithm employed, we reserve this study for a future work.\\\\
Finally, we suggest future improvements and extensions of the algorithm, including its application for detecting single transits, exocomets, flares, and exoplanets featuring transit time variations (TTVs), underscoring its versatility and potential for broader impact in astronomical research.
\\\\
The software presented in this work is open source under the MIT License and is
available at \href{https://github.com/lgrcia/nuance}{https://github.com/lgrcia/nuance}, with documentation and tutorials hosted at \href{https://nuance.readthedocs.io}{https://nuance.readthedocs.io}. 

\newpage
We would like to thank Julien De Wit and Prajwal Niraula for meaningful discussions at the beginning of this project, Germain Garcia for his useful insights about numerical optimization, Michaël Gillon for his overall support, the member of the Exotic group at the University of Liège, and the member of the Astronomical Data Group at the Center for Computational Astrophysics for many enriching discussions and feedback. \review{We would also like to thank the journal referee for their thorough and encouraging report, bringing some very meaningful additions to the paper.}{} This publication benefits from the support of the French Community of Belgium in the context of the FRIA Doctoral Grant awarded to L.J.G. F.J.P acknowledges financial support from grant CEX2021-001131-S funded by MCIN/AEI/ 10.13039/501100011033 and through projects PID2019-109522GB-C52 and PID2022-137241NB-C4. S.A. acknowledges funding from the European Research Council (ERC) under grant agreement 865624 (GPRV) and from the UK Science and Technology Facilities Council (STFC) under grants ST/S000488/1 and ST/R004846/1. This paper includes data collected by the TESS mission. Funding for the TESS mission is provided by NASA's Science Mission Directorate and NASA Explorer Program. We acknowledge the use of public TESS data from pipelines at the TESS Science Office and at the TESS Science Processing Operations Center. Resources supporting this work were provided by the NASA High-End Computing (HEC) Program through the NASA Advanced Supercomputing (NAS) Division at Ames Research Center for the production of the SPOC data products. TESS data were obtained from the MAST data archive at the Space Telescope Science Institute (STScI). STScI is operated by the Association of Universities for Research in Astronomy, Inc., under NASA contract NAS 5–26555. The Center for Computational Astrophysics at the Flatiron Institute is supported by the Simons Foundation.\\\\

\software{{numpy} \citep{harris2020array}, {scipy} \citep{2020SciPy-NMeth}, {matplotlib} \citep{Hunter:2007}, {jax} \citep{jax2018github}, {jaxopt} \citep{jaxopt_implicit_diff}, {tinygp} \citep{foreman_mackey_2024_10463641}, {lightkurve} \citep{2018ascl.soft12013L}, \review{{astropy} (\citealt{astropy:2013,astropy:2018,astropy:2022})}, {pandas} \citep{mckinney-proc-scipy-2010, reback2020pandas}, {wotan} \citep{wotan}, {snakemake} \citep{Snakemake} \review{and nuance version 0.6.0 \citep{zenodo}}.}{}

\newpage
\appendix
\section{SHO kernel}\label{app_gp}
In order to model stellar variability and its effect on transit detection, we employ a simple physically-motivated GP kernel, describing stellar variability through the covariance of a stochastically-driven damped harmonic oscillator (SHO, \citealt{celerite, celerite2}) taking the form 
\begin{equation}
    \begin{gathered}
        k(\tau) = \sigma^2\,\exp\left(-\frac{\omega\,\tau}{2\,Q}\right)
        \left\{\begin{array}{ll}
            1 + \omega\,\tau & \mbox{for } Q = 1/2 \\
            \cosh(f\,\omega\,\tau/2\,Q) + \sinh(f\,\omega\,\tau/2\,Q)/f
                & \mbox{for } Q < 1/2 \\
            \cos(g\,\omega\,\tau/2\,Q) + \sin(g\,\omega\,\tau/2\,Q)/g
                & \mbox{for } Q > 1/2
        \end{array}\right. \\
        \text{where}\quad \tau = |t_i - t_j|\text{,}\quad f = \sqrt{1 - 4\,Q^2} \quad \text{and}\quad g = \sqrt{4\,Q^2 - 1}
    \end{gathered}
\end{equation}
where $Q$ is the quality factor of the oscillator, $\omega$ its pulsation, and $\sigma$ the amplitude of the kernel function. GP computations in this paper use the implementation from \texttt{tinygp}\footnote{\href{https://github.com/dfm/tinygp}{https://github.com/dfm/tinygp}}, a Python package exposing the quasi-separable kernels from \cite{celerite2} and powered by \textsf{JAX}\footnote{\href{https://github.com/google/jax}{https://github.com/google/jax}}.

\section{Proof for the periodic search expression}\label{proof}

\newcommand{\sumTk}{i\neq k}
From the linear search presented in \autoref{linear_search}, we retain and index by $k$ the parameters of the $K$ individual transits whose epochs $\{T_k\}_k$ are compatible with a periodic signal of period $P$ and epoch $T_0$. From the likelihoods of these transits (computed in \autoref{linear_search}), we want an expression for
\begin{equation*}
    p(\bm{f} \vert P, T_0 ,D, \Delta) = \prod_{k\in\mathbb{T}} p(\bm{f} \vert T_k, D, \Delta),
\end{equation*}
i.e., given a depth $D$, the likelihood of the data given a periodic transit signal of period $P$, epoch $T_0$ and a common depth $\Delta$. Since only $\{p(\bm{f} \vert T_k, D, \Delta_k)\}_{k}$ is known (i.e. transits with different depths), we decompose
\begin{equation}\label{eq:non_part_of_per}
    p(\bm{f} \vert T_k, D, \Delta) = \int p(\bm{f} \vert T_k, D, \tilde\Delta)p(\tilde\Delta | \Delta)\, d\tilde\Delta,
\end{equation}
where $p(\bm{f} \vert T_k, D, \tilde\Delta)$ is the probability of the $k$-th transit to have a depth $\tilde\Delta$ and $p(\tilde\Delta | \Delta)$ the probability to observe the depth $\tilde\Delta$ knowing the existence of a common depth $\Delta$. In other words, \autoref{eq:non_part_of_per} involves the likelihood of the nonperiodic transit $k$ to be part of a periodic transit signal with a common depth $\Delta$.
\\\\
Since each depth $\Delta_k$ is found through generalized least squares, each follows a normal distribution $\mathcal{N}(\Delta_k, \sigma_k^2)$, centered on $\Delta_k$ with variance $\sigma_k^2$ and an amplitude $\mathcal{L}_k$, leading to the likelihood function
\begin{equation*}
    p(\bm{f} \vert T_k, D, \tilde\Delta) = \mathcal{L}_k\exp \left(-\frac{(\tilde\Delta-\Delta_k)^2}{2\sigma_k^2}\right).
\end{equation*}
As for the common transit depth $\Delta$, it can be estimated through the joint probability of all other transit depths than $\Delta_k$, such that
\begin{equation*}
    \Delta \sim \prod_{\sumTk}^K \mathcal{N}(\Delta_i, \sigma_i^2),
\end{equation*}
with 
\begin{equation}\label{eq:params}
\frac{1}{\sigma^2} = \sum_{\sumTk}^K \frac{1}{\sigma_i^2} \hspace{0.5cm} \text{and} \hspace{0.5cm}
\Delta =\sigma^2 \sum_{\sumTk}^K {\frac{\Delta_i}{\sigma_i^2}}.
\end{equation}
Hence
\begin{equation*}
    p(\tilde\Delta | \Delta) = \frac{1}{\sqrt{2\pi\sigma^2}}\exp \left(-\frac{(\tilde\Delta-\Delta)^2}{2\sigma^2}\right).
\end{equation*}
We can now rewrite \autoref{eq:non_part_of_per} as
\begin{equation*}
    p(\bm{f} \vert T_k, D, \Delta) =  \frac{\mathcal{L}_k}{\sqrt{2\pi\sigma^2}} \int \exp\left(-\frac{(\tilde\Delta-\Delta_k)^2}{2\sigma_k^2}\right)\, \exp\left(-\frac{(\tilde\Delta-\Delta)^2}{2\sigma^2}\right)\, d\tilde\Delta.
\end{equation*}
The integral in this equation is a product of gaussian integrals that can be obtained analytically, leading to
\begin{equation*}
    p(\bm{f} \vert T_k, D, \Delta) = \mathcal{L}_k  \sqrt{\frac{\sigma_{k}^2}{\sigma^{2} + \sigma_{k}^{2}}} \exp\left(-\frac{1}{2}\frac{(\Delta_k-\Delta)^2}{\sigma_k^2 + \sigma^2}\right).
\end{equation*}
Finally,
\begin{equation}
    \ln p(\bm{f} \vert P, T_0 ,D, \Delta) =  \sum_{k}^K \ln \mathcal{L}_k  - \frac{1}{2} \sum_k^K\left(\ln(\sigma_{k}^2) - \ln(\sigma^{2} + \sigma_{k}^{2}) +  \frac{\left(\Delta_{k} -
    \Delta\right)^{2}}{\sigma_k^{2} + \sigma^{2}}\right),
\end{equation}
the log-likelihood of the data given a periodic transit signal of period $P$, epoch $T_0$, duration $D$ and common depth $\Delta$. In order to reduce the number of times \autoref{eq:params} is computed, we adopt the biased estimates
\begin{equation}\label{eq:sigma_delta}
    \frac{1}{\sigma^2} = \sum_{k}^K \frac{1}{\sigma_i^2} \hspace{0.5cm}\text{and}\hspace{0.5cm} \Delta  = \sigma^2 \sum_{k}^K {\frac{\Delta_i}{\sigma_i^2}},
\end{equation}
so that $\Delta$ and $\sigma$ are independent of $k$ in the last sum of \autoref{eq:result}.

\newpage
\section{Injection-recovery on TIC 1019692}\label{1019692}

\begin{figure}[H]
    \begin{centering}
        \includegraphics[width=\linewidth]{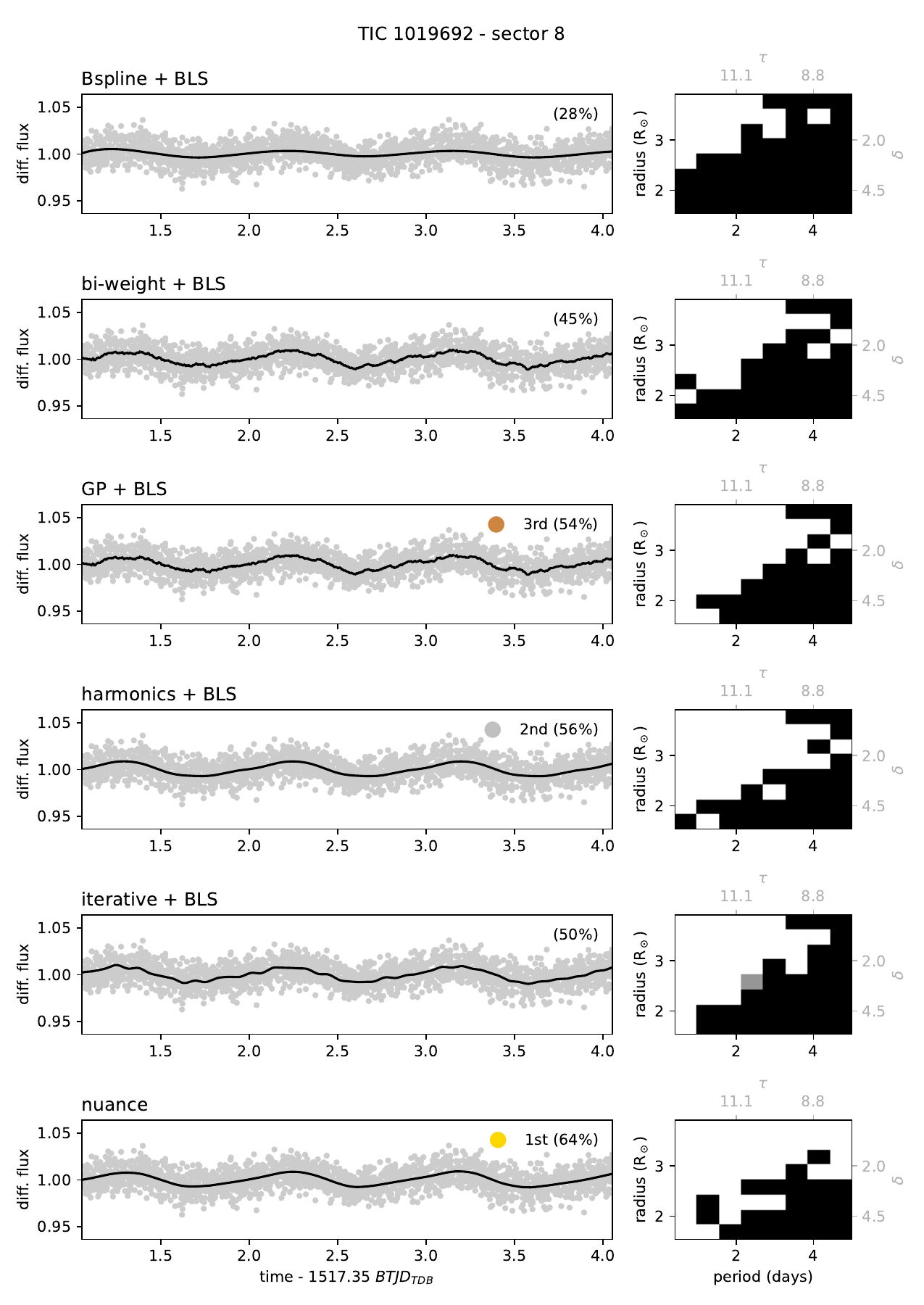}
    \end{centering}
\end{figure}

\begin{figure}[H]
    \begin{centering}
        \caption{Results of the transits injection-recovery on TIC 1019692 half-sector light curve. Left: cleaned light curve with computed trend overplotted in black (except for \textsf{nuance} where it corresponds to the mean of the GP model). Right: Results of the transit search where a black square denotes a transit signal not detected, gray a signal detected at an alias period ($P/2$ or $2P$), and  white a signal detected with the correct period. On the right plots, secondary axes show the $(\tau, \delta)$ relative parameter space. For each method, the upper right legend on the left plot indicates its ranking based on the percent of recovered transit signals (where a transit with an aliased period counts as being detected). \codelink{workflows/tess_injection_recovery/scripts/plot_comparison.py}
        }
        \label{fig:onesearch}
    \end{centering}
\end{figure}

\bibliography{ref}

\end{document}